\documentclass[12pt]{article}
\usepackage{amsmath}
\usepackage{amssymb}
\newsymbol \blackbox 1004
\newcommand{\eh}{\hfill}\newlength{\sperr}

\newenvironment{proof}{{\settowidth{\sperr}{\bf\rm Proof}%
\par\addvspace{0.3cm}\noindent\parbox[t]{1.3\sperr}
{\bf\rm P\eh r\eh o\eh o\eh f\eh }%
}}{\nopagebreak\mbox{} $\blackbox$\par\addvspace{0.3cm}}

\def\a{\alpha}

\def\s{\sigma}
\def\l{\lambda}
\def\L{\Lambda}

\def\t{\theta}

\def\vp{\varphi}
\def\ve{\varepsilon}
\def\wh{\widehat}
\def\wt{\widetilde}

\def\BC{{\mathbb C}}
\def\BR{{\mathbb R}}
\def\clp{{\mathcal P}}
\def\cla{{\mathcal A}}
\def\clb{{\mathcal B}}

\def\cln{{\mathcal N}}
\def\clu{{\mathcal U}}

\newtheorem{Pa}{Paper}[section]
\newtheorem{Tm}[Pa]{{\bf Theorem}}

\newtheorem{Rk}[Pa]{{\bf Remark}}

\newtheorem{Dn}[Pa]{{\bf Definition}}

\newcommand{\CC}
{{\mathchoice {\setbox0=\hbox{$\displaystyle\rm C$}\hbox{\hbox
to0pt{\kern0.4\wd0\vrule height0.9\ht0\hss}\box0}}
{\setbox0=\hbox{$\textstyle\rm C$}\hbox{\hbox
to0pt{\kern0.4\wd0\vrule height0.9\ht0\hss}\box0}}
{\setbox0=\hbox{$\scriptstyle\rm C$}\hbox{\hbox
to0pt{\kern0.4\wd0\vrule height0.9\ht0\hss}\box0}}
{\setbox0=\hbox{$\scriptscriptstyle\rm C$}\hbox{\hbox
to0pt{\kern0.4\wd0\vrule height0.9\ht0\hss}\box0}}}}

\title{Discrete canonical system and non-Abelian Toda lattice:
B\"acklund-Darboux transformation, Weyl functions, and explicit
solutions}

\author{A.L. Sakhnovich}

\date{}
\begin{document}
\maketitle

Branch of Hydroacoustics, Marine Institute of Hydrophysics,  \\
National Academy of Sciences of Ukraine

e-mail addresses: al$_-$sakhnov@yahoo.com

\vspace{+4mm}

{\bf Short title.} Discrete canonical system

\begin{abstract}  A version of  the iterated B\"acklund-Darboux
transformation, where Darboux matrix takes a form of the transfer
matrix function from the system theory,  is constructed for the
discrete canonical system and Non-Abelian Toda lattice.  Results
on the transformations of the Weyl functions, insertion of the
eigenvalues, and construction of the bound states are obtained. A
wide class of the explicit solutions is given. An application to
the semi-infinite block Jacobi matrices is treated.
\end{abstract}

\section{Introduction} \label{intro}

\setcounter{equation}{0}   The matrix discrete canonical system
(DCS) has the form \cite{RS}, \cite{SaL2}:
\begin{equation} \label{1.1}
w(k, \l)-w(k-1, \l)=i \l J H_k w(k-1, \l) \quad (k \geq 1),
\end{equation}
where $H_k$ are $m \times m$  matrices (the set $\{ H_k \}$ we
sometimes call Hamiltonian), and
\begin{equation} \label{1.2}
J=J^*=J^{-1}, \quad H_k=H_k^*, \quad H_k J H_k =0.
\end{equation}
This system is a discretization of the  classical canonical system
\cite{dB, GK, SaL3}:
\[
\frac{d}{d x}w(x, \l) = i \l J H(x) w(x, \l) \quad (H(x) \geq 0).
\]
DCS is equivalent to an important subclass of the well-known
Jacobi systems \cite{RS} and it is an auxiliary system for the
matrix Toda lattice (MTL) studied in \cite{SaL2}:
\begin{equation} \label{3.2}
\displaystyle{\left\{
\begin{array}{c} \displaystyle{Q^{\prime}(k,t)=C(k+1,t)C(k,t)^{-1}-
C(k,t)C(k-1,t)^{-1}} \\
\displaystyle{C^{\prime}(k,t)=Q(k,t)C(k,t).}
\end{array}  \right.}
\end{equation}
Here $C(k,t)$ and $Q(k,t)$ are $p \times p$ ($p=m/2$) matrix
functions, $Q^{\prime}(t)=\frac{d}{dt}Q(t)$, and
\begin{equation} \label{3.3}
 C(k,t)Q(k,t)^*- Q(k,t)C(k,t)=0, \quad C(k,t)=C(k,t)^*, \quad \det
 \, C(k,t) \not= 0.
\end{equation}
If   system (\ref{3.2})  is valid, $C(k,t)>0$ and $p=1$, then the
first relation in (\ref{3.3}) holds automatically and  the
functions $u(k,t)=- \ln \, C(k,t)$ satisfy the classical Toda
lattice
\begin{equation} \label{3.1}
\frac{d^2}{dt^2}u(k,t)= \exp [u(k-1,t)-u(k,t)]-\exp
[u(k,t)-u(k+1,t)].
\end{equation}
 The Non-Abelian Toda lattice (NTL) was introduced
by Polyakov \cite{Kr} as a discretization of the principal chiral
field equation and coincides with system (\ref{3.2}) without
additional condition (\ref{3.3}). When the auxiliary Jacobi
matrices are used NTL is  presented sometimes \cite{G} in the form
\begin{equation} \label{4.1}
\cla^{\prime}_k(t)=\clb_{k+1}(t) \cla_{k}(t)- \cla_{k}(t)
\clb_{k}(t), \quad \clb^{\prime}_k(t)=\cla_{k}(t)- \cla_{k-1}(t)
\quad (k>0)
\end{equation}
The equivalence between systems (\ref{3.2}) and (\ref{4.1}) is
given by the relations
\begin{equation} \label{4.2}
\cla_{k}(t)=C(k+1,t)C(k,t)^{-1}, \quad \clb_k(t)=Q_k(t),
\end{equation}
and vice versa:
\begin{equation} \label{4.3}
Q_k(t)= \clb_k(t), \quad C(k,t)= \left( \prod_{l=0}^{k-1}
\cla_{l}(t) \right) C(0,t),
\end{equation}
\begin{equation} \label{4.4}
C^{\prime}(0,t)=\cla_{0}(t)^{-1}(Q(1,t)-
\cla^{\prime}_{0}(t)\cla_{0}(t)^{-1}) \cla_{0}(t) C(0,t).
\end{equation}
Interesting spectral results for the Toda lattice and its MTL and
NTL generalizations have been obtained in \cite{B, BMRL, BGHT, FG,
 K, KV, SaL2, T, V}  (see also references), though various related
spectral problems are still to be solved.

We shall construct a B\"acklund-Darboux transformation (BDT) for
the DCS (\ref{1.1}), MTL and NTL. The BDTs are widely used in the
spectral and in the integrable nonlinear equations theories, and
the construction of the explicit solutions is one of the important
applications. Numerous results and literature on BDTs are
contained in \cite{AS, CD, GSS, M, MS, Mi, ZM}. In spite of
various important results (see \cite{AS, BR, CC, EJ, KSS, MS, To}
and references therein) the BDTs for the discrete equations are
more complicated and less studied than in the continuous case
\cite{AS}. An interesting modification of the BDT have been
suggested by V.A. Marchenko. Some developments and applications to
the explicit solutions of the Toda lattice and non-Abelian
two-dimensional Toda lattice one can find in \cite{S} and
\cite{EGR}, respectively. The version of the BDT that we are going
to apply (GBDT in the terminology of \cite{SaA4}) was initially
developed in \cite{SaA1,SaA2} (see the case of the canonical
system in \cite{SaA3} and more references and applications in
\cite{GKS}, \cite{SaA4}-\cite{SZ}). The Darboux matrix in this
approach is presented as the transfer matrix function, which is a
well known tool in system theory \cite{KFA}. Transfer matrix
function of the form
\begin{equation} \label{LS}
w_A( \l)=I_m-  \Pi(2)^* S^{-1}(A(1) - \l I_n)^{-1} \Pi(1) \quad
(A(1)S-S A(2)= \Pi(1) \Pi(2)^*),
\end{equation}
that we are going to use, was introduced by L. Sakhnovich in the
context of his method of operator identities \cite{SaL1, SaL2}.
Representation (\ref{LS}) takes roots in the Liv\v{s}ic-Brodskii
characteristic matrix function \cite{L}.

GBDT for the discrete canonical system is constructed in Section
2. The results on the corresponding to GBDT transformations of the
Weyl functions and Theorem \ref{TmIns} on the insertion of the
eigenvalues and construction of the bound states are given in
Section 3. GBDT for the matrix Toda lattice is constructed in
Section 4 and its modification for the Non-Abelian Toda lattice is
discussed in Section 6. A wide class of the explicit solutions of
the Non-Abelian Toda lattice, the corresponding fundamental
solutions of the discrete canonical systems, and evolution of the
Weyl functions are described in Section 5. Appendix is dedicated
to the analog of Theorem \ref{TmIns} for the block Jacobi
matrices.
%%%%%%%%%%%%%%%%%%%%%%%%%%%%%%%%%%%%%%%%%%%%%%%%%%%%%%%%%%%%%
\section{Discrete canonical system: GBDT and Weyl functions}
\label{GBDT} \setcounter{equation}{0} \vspace{-1mm}

To introduce the GBDT of the discrete canonical system suppose
that system (\ref{1.1}), (\ref{1.2}) is given and fix an integer
parameter $n>0$ and three parameter matrices: $n \times n $
matrices $A$ and $S_0$ and $n \times m$ matrix $\Pi_0$ satisfying
the matrix identity
\begin{equation} \label{2.1}
AS_0-S_0 A^*=i \Pi_0 J \Pi_0^* \quad (S_0=S_0^*).
\end{equation}
Then the sets of matrices $\Pi_k$ and $S_k$ are defined by the
equalities
\begin{equation} \label{2.2}
\Pi_k = \Pi_{k-1}- i A  \Pi_{k-1} J H_k \quad (k \geq 1),
\end{equation}
\begin{equation} \label{2.3}
S_k = S_{k-1}+ \Pi_{k-1} J H_k J \Pi_{k-1}^* \quad (k \geq 1).
\end{equation}
System (\ref{2.2}) is obtained from the system dual to (\ref{1.1})
via the substitution of the spectral parameter $\l$ by the
generalized eigenvalue - matrix $A$. System (\ref{2.3}) is chosen
so that the identities
\begin{equation} \label{2.5}
AS_k-S_k A^*=i \Pi_k J \Pi_k^*  \quad (k \geq 0)
\end{equation}
hold. Indeed, in view of (\ref{2.3}) we have
\[
AS_k-S_k A^*=AS_{k-1}-S_{k-1}A^*+ A \Pi_{k-1} J H_k J \Pi_{k-1}^*
- \Pi_{k-1} J H_k J \Pi_{k-1}^* A^*.
\]
Taking into account (\ref{1.2}) and (\ref{2.2}) we rewrite the
equality above:
\begin{equation} \label{2.4}
AS_k-S_k A^*=AS_{k-1}-S_{k-1}A^*+i(\Pi_k J \Pi_k^*-\Pi_{k-1} J
\Pi_{k-1}^*).
\end{equation}
Formulas (\ref{2.1}) and (\ref{2.4}) yield  identities
(\ref{2.5}). Define now a transfer matrix function $w_A(k, \l)$
($\l \not\in \s(A), \, \s$ means spectrum) in the  form
(\ref{LS}):
\begin{equation} \label{2.6}
w_A(k, \l)=I_m-i J \Pi_k^* S_k^{-1}(A- \l I_n)^{-1} \Pi_k.
\end{equation}
\begin{Tm} \label{Tm2.1}
Let matrix function $w$ satisfy initial DCS  (\ref{1.1}),
(\ref{1.2}). Choose integer $n>0$ and  three parameter matrices:
$n \times n$ matrices $A$ and $S_0$ and $n \times m$ matrix
$\Pi_0$ such that (\ref{2.1}) holds. Define matrices $\{ \Pi_k \}$
and $\{ S_k \}$ $(k>0)$ via (\ref{2.2}) and (\ref{2.3}), and
suppose $\det \, S_k \, \not = 0$ $(0 \leq k \leq K)$. Choose a
$J$-unitary $m \times m$ matrix $w_0(0)$, i.e., choose $w_0(0)$
such that $w_0(0)J w_0(0)^*=J$, and define matrices $w_0(k)$ $(0<
k \leq K)$ by the relations
\begin{equation} \label{2.7}
w_0(k)=(I_m- J \Pi_k^* S_k^{-1} \Pi_k J H_k  + J H_k J \Pi_{k-1}^*
S_{k-1}^{-1} \Pi_{k-1}) w_0(k-1).
\end{equation}
Then a solution $\wt w(k, \l)\quad (0 < k  \leq K)$ of the
transformed DCS
\begin{equation} \label{2.9}
{\wt w}(k, \l)- {\wt w}(k-1, \l)=i \l J {\wt H}_k {\wt w}(k-1, \l)
,
\end{equation}
where
\begin{equation} \label{2.10}
{\wt H}_k=w_0(k)^* (H_k + H_k J  \Pi_{k-1}^* S_{k-1}^{-1}
\Pi_{k-1} J H_k )w_0(k),
\end{equation}
\begin{equation} \label{2.11}
{\wt H}_k J {\wt H}_k=0,
\end{equation}
is given by the formulas
\begin{equation} \label{2.8}
{\wt w}(k, \l)=v(k, \l)w(k, \l), \quad v(k, \l)=w_0(k)^{-1}w_A(k,
\l).
\end{equation}
Moreover if $H_k \geq 0$ $( 0< k \leq K)$ and $S_0>0$, then
$S_k>0$ and ${\wt H}_k \geq 0$ also.
\end{Tm}
System (\ref{2.9}) is the GBDT of the initial system (\ref{1.1}).
Its Hamiltonian ${\wt H_k}$ and fundamental solution are expressed
explicitly in terms of the Hamiltonian and fundamental solution of
the initial system (see formulas (\ref{2.10}) and (\ref{2.8})),
which is the characteristic property of the BDT. GBDT (\ref{2.9})
is determined by the parameter matrices $A$, $S_0$, and $\Pi_0$,
satisfying identity (\ref{2.1}).
\begin{proof} of Theorem \ref{Tm2.1}.
From (\ref{2.3}) it follows easily that
\begin{equation} \label{2.12}
S_k^{-1}-S_{k-1}^{-1}=-S_k^{-1} \Pi_{k-1}J H_k J \Pi_{k-1}^*
S_{k-1}^{-1}.
\end{equation}
In view of (\ref{2.12}) we get
\[
\Pi_k^* S_k^{-1}- \Pi_{k-1}^* S_{k-1}^{-1}= \Pi_{k}^*
S_{k-1}^{-1}- \Pi_{k}^*S_k^{-1} \Pi_{k-1}J H_k J \Pi_{k-1}^*
S_{k-1}^{-1}- \Pi_{k-1}^* S_{k-1}^{-1}.
\]
Using (\ref{2.2}) we rewrite this equality as
\begin{equation} \label{2.13}
\Pi_k^* S_k^{-1}- \Pi_{k-1}^* S_{k-1}^{-1}= i H_k J \Pi_{k-1}^*A^*
S_{k-1}^{-1}- \Pi_{k}^*S_k^{-1} \Pi_{k-1}J H_k J \Pi_{k-1}^*
S_{k-1}^{-1}.
\end{equation}
By (\ref{2.5}) we have $A^*S_{k-1}^{-1}=S_{k-1}^{-1}A-i
S_{k-1}^{-1} \Pi_{k-1} J \Pi_{k-1}^* S_{k-1}^{-1}$ and so equality
(\ref{2.13}) takes the form
\[
\Pi_k^* S_k^{-1}- \Pi_{k-1}^* S_{k-1}^{-1}= i H_k J \Pi_{k-1}^*
S_{k-1}^{-1}A+
\]
\begin{equation} \label{2.14}
+(H_k J \Pi_{k-1}^*S_{k-1}^{-1} \Pi_{k-1} J- \Pi_{k}^*S_k^{-1}
\Pi_{k-1}J H_k J) \Pi_{k-1}^* S_{k-1}^{-1}.
\end{equation}
Now we can derive the crucial equality
\[
Z(\l):=w_A(k, \l)(I_m+ i \l J H_k)=
\]
\begin{equation} \label{2.15}
(I_m+ i \l J H_k-J \Pi_k^*S_k^{-1} \Pi_k J H_k+J H_k J
\Pi_{k-1}^*S_{k-1}^{-1} \Pi_{k-1})w_A(k-1, \l).
\end{equation}
Indeed, from (\ref{2.6}), (\ref{2.2}) and (\ref{1.2}) it follows
that
\[
Z(\l)= (I_m-i J \Pi_k^* S_k^{-1}(A- \l I_n)^{-1} \Pi_{k-1})(I_m+ i
\l J H_k)
\]
\begin{equation} \label{2.16}
- J \Pi_k^* S_k^{-1}(A- \l I_n)^{-1} A \Pi_{k-1} J H_k.
\end{equation}
Rewriting $A$ as $(A- \l I_n) + \l I_n$ we obtain
\begin{equation} \label{2.17}
Z(\l)=I_m+ i \l J H_k -i J \Pi_k^* S_k^{-1}(A- \l I_n)^{-1}
\Pi_{k-1}- J \Pi_k^* S_k^{-1} \Pi_{k-1} J H_k.
\end{equation}
In view of (\ref{2.14}) equality (\ref{2.17}) yields
\[
Z(\l)= i \l J H_k + w_A(k-1, \l)+ J  H_k J \Pi_{k-1}^*
S_{k-1}^{-1}A(A- \l I_n)^{-1} \Pi_{k-1}
\] \[
-i J(H_k J \Pi_{k-1}^*S_{k-1}^{-1} \Pi_{k-1} J- \Pi_{k}^*S_k^{-1}
\Pi_{k-1}J H_k J) \Pi_{k-1}^* S_{k-1}^{-1} (A- \l I_n)^{-1}
\Pi_{k-1}
\]
\begin{equation} \label{2.18}
 -J
\Pi_k^* S_k^{-1} \Pi_{k-1} J H_k.
\end{equation}
By definition (\ref{2.6}) and substitution $A=(A- \l I_n) + \l
I_n$ we derive now
\begin{equation} \label{2.19}
Z(\l)= (I_m+i \l J H_k -J \Pi_k^* S_k^{-1} \Pi_{k-1} J H_k+ J H_k
J \Pi_{k-1}^* S_{k-1}^{-1} \Pi_{k-1})w_A(k-1, \l).
\end{equation}
By (\ref{1.2}), (\ref{2.2}) and (\ref{2.19}) formula (\ref{2.15})
is immediate.

Now in view of (\ref{1.1}) and (\ref{2.15}) one gets
\[
w_A(k, \l)w(k, \l)=(I_m+ i \l J H_k-J \Pi_k^*S_k^{-1} \Pi_k J
H_k+J H_k J \Pi_{k-1}^*S_{k-1}^{-1} \Pi_{k-1}) \times
\]
\begin{equation} \label{2.24}
\times w_A(k-1, \l)w(k-1, \l).
\end{equation}
According to (\ref{2.24}) to prove that $\wt w$ of the form
(\ref{2.8}) satisfies (\ref{2.9}) it remains to show that
\begin{equation} \label{2.24.1}
I_m+ i \l J {\wt H}_k=
\end{equation}
\[
w_0(k)^{-1}(I_m+i \l J H_k -J \Pi_k^* S_k^{-1} \Pi_{k-1} J H_k+ J
H_k J \Pi_{k-1}^* S_{k-1}^{-1} \Pi_{k-1}) w_0(k-1).
\]
By the definition (\ref{2.7}) formula (\ref{2.24.1}) is equivalent
to the relation
\begin{equation} \label{2.24'}
J {\wt H}_k=w_0(k)^{-1} J H_k w_0(k-1).
\end{equation}

It follows  \cite{SaL1, SaL2} from (\ref{2.5}) and (\ref{2.6})
that
\begin{equation} \label{2.20}
w_A(k, \l)J w_A(k, \overline{\l})^*=J \quad (k \geq 0).
\end{equation}
By (\ref{1.2})  it is immediate that
\begin{equation} \label{2.21}
(I_m+ i \l J H_k)J(I_m+ i \l J H_k)^*=J \quad (\l \in \BR).
\end{equation}
According to (\ref{2.15}), (\ref{2.20}) and (\ref{2.21}) matrices
\[
I_m-J \Pi_k^*S_k^{-1} \Pi_k J H_k+J H_k J \Pi_{k-1}^*S_{k-1}^{-1}
\Pi_{k-1}
\]
are $J$-unitary. Hence, taking into account formula (\ref{2.7}),
we obtain
\begin{equation} \label{2.22}
w_0(k)J w_0(k)^*=J,
\end{equation}
\begin{equation} \label{2.22.1}
w_0(k-1)=J(I_m- J \Pi_k^* S_k^{-1} \Pi_k J H_k + J H_k J
\Pi_{k-1}^* S_{k-1}^{-1} \Pi_{k-1})^*J w_0(k).
\end{equation}
Finally in view of the  formulas (\ref{1.2}), (\ref{2.7}),
(\ref{2.10}), and (\ref{2.22}) we derive
\begin{equation} \label{2.23}
J {\wt H}_k= J w_0(k)^* (H_k + H_k J  \Pi_{k-1}^* S_{k-1}^{-1}
\Pi_{k-1} J H_k )w_0(k)=
\end{equation}
\[
w_0(k)^{-1}J H_k J(I_m+ \Pi_{k-1}^* S_{k-1}^{-1} \Pi_{k-1}J H_k J
- H_k J \Pi_k^* S_k^{-1} \Pi_k J )J w_0(k).
\]
In view of (\ref{2.22.1}) and (\ref{2.23}) the equality
(\ref{2.24'}) is true, and so (\ref{2.9}) is satisfied. Equalities
(\ref{2.11}) follow from (\ref{2.10}) and (\ref{2.22}). The last
statement of the theorem is immediate from (\ref{2.3}) and
(\ref{2.10}).
\end{proof}

\begin{Rk} \label{Rk2.2}
When $A$ is invertible and $w_0(0)=w_A(0,0)$ then by (\ref{2.7})
and (\ref{2.15}) we have $w_0(k)=w_A(k,0)$ for all $k \geq 0$.
\end{Rk}
\begin{Rk} \label{Rk2.3} Formula (\ref{2.20}) is a particular case
of a more general relation that follows  \cite{SaL1, SaL2} from
(\ref{2.5}) and (\ref{2.6}):
\begin{equation} \label{2.25}
w_A(k, \l)^*J w_A(k, \mu)=J+ i(\overline{\l}- \mu) \Pi_k^*(A^*-
\overline{\l}I_n)^{-1}S_k^{-1}(A- \mu I_n)^{-1} \Pi_k.
\end{equation}
\end{Rk}
\section{Transformation of the Weyl functions and insertion
of the eigenvalues} \label{Wtr}
\setcounter{equation}{0} Let now $m=2p$, $H_k \geq 0$ $(0< k \leq
K )$, and put
\begin{equation} \label{2.28}
J=\left[
\begin{array}{lr}
0 & I_p \\ I_p & 0
\end{array}
\right].
\end{equation}
Normalize the $m \times m$  fundamental solution $w$ of DCS
 by the initial condition
\begin{equation} \label{2.30}
w(0, \l)=I_m.
\end{equation}
\begin{Dn} \label{Dnj1} \cite{SaL2}
The set  $\cln (w,K)$ of the Weyl functions $\vp(\l)$ for DCS
(\ref{1.1}) given on the interval $0<k \leq K$ is defined
 by the linear fractional transformation
\begin{equation} \label{2.28.1}
\vp(\l)=i[I_p \quad 0]w(K, {\overline{\l}})^* \chi(\l) \Big( [0
\quad I_p]w(K, {\overline{\l}})^* \chi(\l) \Big)^{-1} \quad (\l
\in \BC_+),
\end{equation}
where $\BC_+$ is the open upper half-plane, and $\chi(\l)$ are
meromorphic in $\BC_+$ and non-degenerate $m \times p$ matrix
functions with the $J$-property:
\begin{equation} \label{2.28.2}
\chi(\l)^* \chi(\l)>0, \quad \chi(\l)^* J \chi(\l) \geq 0.
\end{equation}
\end{Dn}
Notice \cite{SaL2} that the inequality
\begin{equation} \label{2.28.3}
\clu(\l):=w(K, {\overline{\l}})^*J w(K, {\overline{\l}}) \geq J
\end{equation}
holds, and if $\det   \Big( [0 \quad I_p]w(K, {\overline{\l}})^*
\chi(\l) \Big) \not= 0$, then $i(\vp - \vp^*) \leq 0$, i.e.,
\begin{equation} \label{2.28.4}
\wh \chi (\l)^* J \wh \chi (\l) \geq 0, \quad \wh \chi
(\l):=\left[
\begin{array}{c}
 - i \vp(\l) \\ I_p
\end{array}
\right].
\end{equation}
If $H_k \geq 0$ for $0<k < \infty$ and $\clu(\l) >J$, then $\det
\, \Big( [0 \quad I_p]w(K, {\overline{\l}})^* \chi(\l) \Big) \not=
0$ for all $\chi$ satisfying (\ref{2.28.2}), and it was shown in
\cite{SaL2} that the Weyl functions $\vp \in \cap_{K< \infty}
\cln(w,K)$ have the characteristic property:
\begin{equation} \label{2.29}
\displaystyle{ \sum_{k=0}^{\infty} \left[
\begin{array}{lr}
I_p & i \vp(\l)^*
\end{array}
\right] w(k, \l)^*H_{k+1} w(k, \l)
 \left[
\begin{array}{c}
I_p \\ - i \vp(\l)
\end{array}
\right]< \, \infty .}
\end{equation}
\begin{Dn} \label{Dnj2}
The  Weyl functions $\vp(\l)$ for DCS (\ref{1.1}) $\,(H_k \geq 0)$
given on the semiaxis $0<k < \infty $ are defined
 by the condition (\ref{2.29}).
 \end{Dn}
In view of the representation (\ref{2.8}) the normalized by the
condition $\wt w(0, \l)=I_m$  solution $\wt w$ of the transformed
system  (\ref{2.9}) is given by the equality
\begin{equation} \label{2.31}
\wt w(k, \l)=v(k, \l)w(k, \l) v(0, \l)^{-1}.
\end{equation}
The analog of Theorem 4 \cite{SaA3} on the Weyl functions of the
transformed canonical system is true.
\begin{Tm} \label{Tm2.4} Suppose the Hamiltonian
of the DCS (\ref{1.1}), (\ref{1.2}) on the interval $0<k \leq K$
is non-negative: $H_k \geq 0$, and the inequality in
(\ref{2.28.3}) is strict: $\clu(\l)
>J$. If $S_0>0$, then the function $\wt \vp$ of the form
\begin{equation} \label{2.32}
\wt \vp(\l)=i[I_p \quad 0]J v(0, \l)J \wh \chi(\l) \Big( [0 \quad
I_p]J v(0, \l)J \wh \chi(\l) \Big)^{-1},
\end{equation}
where   $\vp \in \cln(w,K)$ and $\wh \chi$ is given by
(\ref{2.28.4}), is a Weyl function of the transformed DCS
(\ref{2.9}), i.e., $\wt \vp \in \cln( \wt w,K)$.
\end{Tm}
\begin{proof}.
Notice  that $J v(0, \l)J=\big( v(0, {\overline{\l}})^*
\big)^{-1}$. Hence according to (\ref{2.28.1}),  (\ref{2.31}), and
definition of $\wh \chi$ in (\ref{2.28.4}) we have
\[
J v(0, \l)J \wh \chi(\l)=\big( v(0, {\overline{\l}})^* \big)^{-1}
 w(K, {\overline{\l}})^* v(K, {\overline{\l}})^* \wt \chi(\l)
 \Big( [0
\quad I_p]w(K, {\overline{\l}})^* \chi(\l) \Big)^{-1}
\]
\begin{equation} \label{2.34}
= \wt w(K, {\overline{\l}})^* \wt \chi(\l)
 \Big( [0
\quad I_p]w(K, {\overline{\l}})^* \chi(\l) \Big)^{-1},
\end{equation}
where $\wt \chi(\l):= \big( v(K, {\overline{\l}})^* \big)^{-1}
\chi(\l)$. By Theorem \ref{Tm2.1} from $S_0>0$ we get $S_k>0$.
Thus formulas (\ref{2.22}) and (\ref{2.25}) yield
\begin{equation} \label{2.33.1}
v(k, \l)^*J v(k, \l) \geq J, \quad v(k, {\overline{\l}})^*J v(k,
{\overline{\l}}) \leq J \quad (\l \in \BC_+).
\end{equation}
In particular, we have
\begin{equation} \label{2.33}
v(k, {\overline{\l}})^{-1}J \big( v(k, {\overline{\l}})^*
\big)^{-1} \geq J \quad (k \geq 0).
\end{equation}
Therefore if $\chi$ satisfies relations (\ref{2.28.2}), then the
matrix function $\wt \chi(\l)= \big( v(K, {\overline{\l}})^*
\big)^{-1} \chi(\l)$ satisfies relations (\ref{2.28.2}) also.
Moreover as $\clu (\l) >J$ we derive from (\ref{2.33}) that
\begin{equation} \label{2.34'}
\big( v(0, {\overline{\l}})^* \big)^{-1}
 w(K, {\overline{\l}})^*J w(K, {\overline{\l}})v(0,
 {\overline{\l}})^{-1}>J.
\end{equation}
Taking into account formulas (\ref{2.28.2}) and (\ref{2.34'})  we
obtain
\begin{equation} \label{2.35}
\det \, \Big( [0 \quad I_p]
 \wt w(K, {\overline{\l}})^* \wt \chi(\l) \Big)=
\det \, \Big( [0 \quad I_p] \big( v(0, {\overline{\l}})^*
\big)^{-1}
 w(K, {\overline{\l}})^* \chi(\l) \Big) \not= 0.
\end{equation}
From (\ref{2.34}), (\ref{2.33}), and (\ref{2.35}) follows the
statement of the theorem.
\end{proof}
The next theorem is closely related to Theorem \ref{Tm2.4}.
\begin{Tm} \label{Tm2.5}
Suppose $H_k \geq 0$ for $0<k < \infty$, $S_0>0$, and  $\vp(\l)$
is a Weyl function for DCS (\ref{1.1}) on the semiaxis $0<k<
\infty$, i.e., matrix function $\vp(\l)$ satisfies (\ref{2.29}).
If  the inequality $\det \Big( [0 \quad I_p]J v(0, \l)J \wh
\chi(\l) \Big) \not=0$ holds, then the matrix function $\wt
\vp(\l)$ given by the equality (\ref{2.32}) is a Weyl function of
the transformed  DCS (\ref{2.9}) on the semiaxis:
\begin{equation} \label{2.36}
\displaystyle{ \sum_{k=0}^{\infty} \left[
\begin{array}{lr}
I_p & i \wt \vp(\l)^*
\end{array}
\right]\wt w(k, \l)^* \wt H_{k+1} \wt w(k, \l)
 \left[
\begin{array}{c}
I_p \\ - i \wt \vp(\l)
\end{array}
\right]< \, \infty }.
\end{equation}
Here $\wt w$ is the  solution of (\ref{2.9}) normalized by the
condition $\wt w(0, \l)=I_m$.
\end{Tm}
\begin{proof}.
From the equality
\[
(I_m+ i \l J H_{k+1})^*J(I_m+ i \l J H_{k+1})=J+i (\l -
\overline{\l})H_{k+1}
\]
and (\ref{1.1}) it follows that
\begin{equation} \label{2.26}
w(k+1, \l)^*J w(k+1, \l)-w(k, \l)^*J w(k, \l)=i (\l -
\overline{\l})w(k, \l)^*H_{k+1} w(k, \l).
\end{equation}
Hence one gets \cite{SaL2}
\begin{equation} \label{2.27}
\displaystyle{ \sum_{k=0}^{K-1}w(k, \l)^*H_{k+1} w(k, \l)=
\frac{i}{\l - \overline{\l}} \big(J-w(K, \l)^*J w(K, \l) \big).}
\end{equation}
In view of (\ref{2.27}) inequality (\ref{2.29}) is equivalent to
the boundedness of the forms
\begin{equation} \label{2.37}
\wh \chi(\l)^*w(K, \l)^*(-J) w(K, \l) \wh \chi (\l)<M(\l)I_m \quad
(K< \infty).
\end{equation}
By the formulas (\ref{2.33.1}) and (\ref{2.37}) we have
\begin{equation} \label{2.38}
\wh \chi(\l)^*w(K, \l)^*v(K, \l)^*(-J)v(K, \l) w(K, \l) \wh \chi
(\l)<M(\l)I_m .
\end{equation}
Formula (\ref{2.32}) yields the equality
\begin{equation} \label{2.39}
\left[
\begin{array}{c}
 - i \wt \vp(\l) \\ I_p
\end{array}
\right]=J v(0, \l)J \wh \chi(\l)\Big( [0 \quad I_p]J v(0, \l)J \wh
\chi(\l) \Big)^{-1}.
\end{equation}
According to relations (\ref{2.31}), (\ref{2.38}) and (\ref{2.39})
the inequalities
\begin{equation} \label{2.40}
[I_p \quad i \wt \vp(\l)^*] \wt w(K, \l)^*(-J) \wt w(K, \l) \left[
\begin{array}{c}
I_p \\ - i \wt \vp(\l)
\end{array}
\right] <M_1(\l)I_m \quad (K< \infty)
\end{equation}
are true. Similar to the equivalence of  (\ref{2.29}) and
(\ref{2.37}) we get (\ref{2.36}) from (\ref{2.40}).
\end{proof}

Insertion of the eigenvalues by the commutation methods have been
studied in the important papers \cite{D, GSS}, and useful
references on the insertion of the eigenvalues into Jacobi
operators one can find in \cite{BGHT, T}. Insertion of the
eigenvalues and construction of the bound states using GBDT
approach were investigated in the continuous case in \cite{GKS0,
SaA3}. The bound states and eigenvalues generated in DCS by the
parameter matrix $A$ are described in the following theorem.
\begin{Tm} \label{TmIns}
Suppose $H_k \geq 0$ for $0<k < \infty$, $S_0>0$ and GBDT
(\ref{2.9}) is determined by the parameter matrices $A$, $S_0$,
and $\Pi_0$. If $f \not=0$ is an eigenvector of $A$, that is $A f=
\mu f$,  then the sequence $\{g_k \}$, where
\begin{equation} \label{j1}
g_k= w_0(k)^{-1}J \Pi_k^*S_k^{-1}f,
\end{equation}
is a bound state of the DCS corresponding to the eigenvalue $\mu$,
i.e., the relations
\begin{equation} \label{j2}
g_k=(I_m+i \mu J {\wt H}_k)g_{k-1} \quad (k>0), \quad
\sum_{k=0}^{\infty}g_k^*{\wt H}_{k+1}g_k < \infty
\end{equation}
are valid.
\end{Tm}
\begin{proof}. First we shall prove the first relation in
(\ref{j2}). By (\ref{1.2}), (\ref{2.2}), and (\ref{2.14}) we get
\[ J \Pi_k^*S_k^{-1}f= \]
\begin{equation} \label{j4}
=(I_m+i \mu J H_k+JH_kJ \Pi_{k-1}^*S_{k-1}^{-1} \Pi_{k-1}-J
\Pi_{k}^*S_{k}^{-1} \Pi_{k}J H_k)J \Pi_{k-1}^*S_{k-1}^{-1}f.
\end{equation}
According to (\ref{2.24.1}), (\ref{j1}), and (\ref{j4}) the first
relation in (\ref{j2}) is true.

To prove the second relation in (\ref{j2}) for $\mu \not=
\overline{\mu}$ analogously to (\ref{2.27}) we get
\begin{equation} \label{j3}
\displaystyle{ \sum_{k=0}^{K-1} \wt w(k, \l)^* \wt H_{k+1} \wt
w(k, \l)= \frac{i}{\l - \overline{\l}} \big(J- \wt w(K, \l)^*J \wt
w(K, \l) \big).}
\end{equation}
From the first relation in (\ref{j2}) it follows that
\begin{equation} \label{j5}
g_k= \wt w(k, \mu)g_0.
\end{equation}
Therefore, if $\mu \not= \overline{\mu}$, then in view of
(\ref{j3}) and (\ref{j5}) we obtain
\begin{equation} \label{j6}
\sum_{k=0}^{K-1}g_k^*{\wt H}_{k+1}g_k=\frac{i}{\mu -
\overline{\mu}} \big(g_0^* J g_0- g_K^*J g_K  \big).
\end{equation}
Finally notice that formulas (\ref{2.5}), (\ref{2.22}), and
(\ref{j1}) yield:
\begin{equation} \label{j7}
g_k^*J g_k=-i f^*(S_k^{-1}A-A^*S_k^{-1})f=-i(\mu - \overline{\mu})
f^*S_k^{-1}f.
\end{equation}
According to Theorem \ref{Tm2.1} the inequalities $S_k>0$ hold.
Hence from (\ref{j6}) and (\ref{j7}) it follows that
\begin{equation} \label{j8}
\sum_{k=0}^{\infty}g_k^*{\wt H}_{k+1}g_k  \leq f^*S_0^{-1}f,
\end{equation}
and the second relation in (\ref{j2}) is valid. The proof of
(\ref{j8}) in the general case that includes  real eigenvalues of
$A$ is somewhat more complicated. For that purpose we shall
consider $g_k^* \wt H_{k+1} g_k$. From (\ref{2.24'}),
(\ref{2.22}), and (\ref{j1}) we derive
\begin{equation} \label{j9}
g_k^* \wt H_{k+1} g_k=f^*S_k^{-1} \Pi_k J \big(
w_0(k+1)w_0(k)^{-1} \big)^*H_{k+1}J \Pi_k^* S_k^{-1}f.
\end{equation}
Thus taking into account  (\ref{1.2}) and (\ref{2.7}) we get
\[ g_k^* \wt H_{k+1} g_k=
\]
\begin{equation} \label{j10}
=f^*S_k^{-1}( \Pi_k J H_{k+1}J \Pi_k^*- \Pi_k J H_{k+1}J
\Pi_{k+1}^*S_{k+1}^{-1} \Pi_{k+1}J H_{k+1}J \Pi_k^* )S_k^{-1}f.
\end{equation}
Using (\ref{1.2}) and (\ref{2.2}) rewrite (\ref{j10}) as
\begin{equation} \label{j11}
g_k^* \wt H_{k+1} g_k=f^*S_k^{-1}( \Pi_k J H_{k+1}J \Pi_k^*- \Pi_k
J H_{k+1}J \Pi_{k}^*S_{k+1}^{-1} \Pi_{k}J H_{k+1}J \Pi_k^*
)S_k^{-1}f.
\end{equation}
Therefore in view of (\ref{2.3}) we obtain
\[g_k^* \wt H_{k+1} g_k=
f^*S_k^{-1}\big(
S_{k+1}-S_k-(S_{k+1}-S_k)S_{k+1}^{-1}(S_{k+1}-S_k)\big)S_k^{-1}f=
 \]
\begin{equation} \label{j12}
=
f^*(S_k^{-1}-S_{k+1}^{-1})f.
\end{equation}
By (\ref{j12}) inequality (\ref{j8}) is true for $\mu \in \BR$
also.
\end{proof}

\section{Matrix Toda lattice }
\label{TC} \setcounter{equation}{0} \vspace{-1mm}

In this section we shall consider  the MTL generalization
(\ref{3.2}), (\ref{3.3}) of the well known Toda lattice. For this
purpose we shall need an equivalent  representation of (\ref{3.2})
in the zero curvature form \cite{SaL2}
\begin{equation} \label{3.39}
\frac{d}{d t}G(k,t, \l)+G(k,t, \l)F(k-1,t, \l)-F(k,t, \l)G(k,t,
\l)=0,
\end{equation}
where
\begin{equation} \label{3.39'}
G(k,t, \l):=- \l \clp_1 +i \xi(k,t), \quad  F(k,t, \l):=- \l
\clp_2 -i  \psi(k,t),
\end{equation}
\begin{equation} \label{3.39''}
{\clp }_1:= \left[
\begin{array}{lr}
I_p & 0 \\ 0 & 0
\end{array}
\right], \quad {\clp }_2:= \left[
\begin{array}{lr}
0 & 0 \\ 0 & I_p
\end{array}
\right], \quad \clp_1 \clp_2= \clp_2 \clp_1=0,
\end{equation}
\begin{equation} \label{3.6}
\xi(k,t):=\left[
\begin{array}{lr}
-i Q(k,t) & C(k,t) \\ C(k,t)^{-1} & 0
\end{array}
\right], \quad \psi(k,t):=\left[
\begin{array}{lr}
0 & C(k+1,t) \\ C(k,t)^{-1} & 0
\end{array}
\right].
\end{equation}
See \cite{AS, CG, FT, FG, LRag} for the references on the lattice
models and \cite{AL} for the discrete version of the zero
curvature equation.

To describe a B\"acklund-Darboux transformation for the MTL
introduce Hamiltonians $\{ H_k(t) \}$ by the equalities
\begin{equation} \label{3.4}
H_k(t):=J U(k,t) \eta(k,t) U(k,t)^* J, \quad \eta(k,t):= \left[
\begin{array}{lr}
C(k,t) & 0 \\ 0 & 0
\end{array}
\right],
\end{equation}
where $J$ is given by the equality (\ref{2.28}),
\begin{equation} \label{3.5}
U(k,t)=U(k-1,t) \xi(k,t)^{-1}, \quad U(0,t)J U(0,t)^*=J.
\end{equation}

Notice that according to formulas (\ref{3.3}), (\ref{3.6}) and
(\ref{3.5}) we have
\begin{equation} \label{3.7}
\xi (k,t)J \xi (k,t)^*=J, \quad U(k,t)J U(k,t)^*=J.
\end{equation}
By (\ref{3.4}) and the second relation in (\ref{3.7}) $H_k$
satisfies condition (\ref{1.2}).

Add now a new variable $t$ in the notations of Section \ref{GBDT}.
Similar to Section 2, if given a solution $\{ C(k,t), \, Q(k,t)
\}$ of (\ref{3.2}), (\ref{3.3}), we choose $n>0$ and three
parameter matrices $A, \, S_0(0)$, and $\wh \Pi_0(0)$ that satisfy
the matrix identity
\begin{equation} \label{3.12}
AS_0(0)-S_0(0)A^*=i \wh \Pi_{0}(0)J \wh \Pi_{0}(0)^* \quad
(S_0(0)=S_0(0)^*)
\end{equation}
to construct a new solution. We put
\begin{equation} \label{3.8}
\wh \Pi_k(t)= \Pi_k(t)U(k,t),
\end{equation}
and so according to (\ref{3.7})  identity (\ref{3.12}) is
equivalent to the identity $AS_0(0)-S_0(0)A^*=i  \Pi_{0}(0)J
\Pi_{0}(0)^*$. In view of (\ref{3.5}), (\ref{3.7}) and (\ref{3.8})
formulas (\ref{2.2}) are equivalent to the equalities
\begin{equation} \label{3.9}
\wh \Pi_k(t)= \wh \Pi_{k-1}(t) \xi(k,t)^{-1}-i A  \wh \Pi_{k-1}(t)
{\clp }_2.
\end{equation}
 Thus $\{ \wh \Pi_k(t) \}$ is uniquely defined via (\ref{3.9}) by the
initial value $\wh \Pi_0(0)$ and equation
\begin{equation} \label{3.10}
\frac{d}{d t} \wh \Pi_0(t)= A  \wh \Pi_{0}(t) {\clp }_2 +i \wh
\Pi_0(t) \psi(0,t).
\end{equation}
Using (\ref{3.4}), (\ref{3.7}) and (\ref{3.8}) rewrite now
equation (\ref{2.3}) in the form
\begin{equation} \label{3.25}
S_k(t)=S_{k-1}(t)+ \wh \Pi_{k-1}(t) \zeta(k,t)  \wh
\Pi_{k-1}(t)^*,
\end{equation}
where $\zeta(k,t)=\xi(k,t)^{-1} \eta(k,t)(\xi(k,t)^*)^{-1}$. By
(\ref{3.6}) we have
\begin{equation} \label{3.20}
 \xi(l,t)^{-1}= \left[
\begin{array}{lr}
0  &  C(l,t)
\\ C(l,t)^{-1}  & i C(l,t)^{-1}Q(l,t)
C(l,t)
\end{array}
\right],
\end{equation}
For the sake of brevity we shall sometimes omit $t$ in our
notations. In view of (\ref{3.2}), (\ref{3.3}), and (\ref{3.20})
it is immediate that
\begin{equation} \label{3.27}
\zeta(k)=\left[
\begin{array}{lr}
0 & 0 \\ 0 & C(k)^{-1}
\end{array}
\right], \quad \frac{d}{d t}\zeta(k) =- \frac{1}{2}\left[
\begin{array}{lr}
0 & 0 \\ 0 & C(k)^{-1}Q(k)+Q(k)^*C(k)^{-1}
\end{array}
\right].
\end{equation}
We define $\{ S_k(t) \}$ by the formulas (\ref{3.25}), initial
value $S_0(0)$, and equation
\begin{equation} \label{3.11}
\frac{d}{d t} S_0(t)=\frac{1}{2} \Big(AS_0(t)+S_0(t)A^*+ i \wh
\Pi_{0}(t) \left[
\begin{array}{lr}
0 & -I_p \\ I_p & 0
\end{array}
\right] \wh \Pi_{0}(t)^* \Big).
\end{equation}
Then the identity
\begin{equation} \label{3.13}
AS_0(t)-S_0(t)A^*=i \wh \Pi_{0}(t)J \wh \Pi_{0}(t)^*=i \Pi_{0}(t)J
 \Pi_{0}(t)^*.
\end{equation}
is valid. Indeed, by formulas (\ref{3.12}), (\ref{3.10}), and
(\ref{3.11})
 for $Z(t)=AS_0(t)-S_0(t)A^*-i \wh \Pi_{0}(t)J \wh
\Pi_{0}(t)^*$ we have $\displaystyle{Z^{\prime}=
\frac{1}{2}(AZ+ZA^*)}$ and $Z(0)=0$. Hence the identity $Z(t)
\equiv 0$ is valid, i.e., (\ref{3.13}) is true. By the
considerations of Section \ref{GBDT} formula (\ref{3.13}) yields
\begin{equation} \label{3.14}
AS_k(t)-S_k(t)A^*=i \Pi_{k}(t)J
 \Pi_{k}(t)^*.
\end{equation}
Formulas (\ref{3.9})-(\ref{3.25}), (\ref{3.11})  introduce $\wh
\Pi_k$ and $S_k$ independently of the considerations of Section 1,
relations (\ref{3.4}), (\ref{3.5}), and (\ref{3.8}) granting
transfer to the auxiliary discrete canonical systems.

Now a B\"acklund-Darboux transformation for MTL is given by the
relations
\begin{equation} \label{3.15}
\wt C(k,t)=C(k,t)+X_{22}(k-1,t) \quad (k>0), \quad \wt C(0,t)=
\big( C(0,t)-X_{11}(0,t) \big)^{-1},
\end{equation}
\begin{equation} \label{3.16}
\wt Q(k,t)=Q(k,t)+i(X_{21}(k-1,t)-X_{21}(k,t))  \quad (k>0),
\end{equation}
where $X_{ij}(k,t)$ are $p \times p$ blocks of the matrix
\begin{equation} \label{3.17}
X(k,t)=\left[
\begin{array}{lr}
X_{11}(k,t) & X_{12}(k,t) \\ X_{21}(k,t) & X_{22}(k,t)
\end{array}
\right]=\wh \Pi_k(t)^* S_k(t)^{-1} \wh \Pi_k(t).
\end{equation}
\begin{Tm} \label{Tm3.1}
Suppose matrix functions $C(k,t)$ and $Q(k,t)$ satisfy MTL
(\ref{3.2}), (\ref{3.3}) and $C(0,t)$ is continuous. Suppose
additionally that matrix functions $C(0,t)^{-1}- X_{11}(0,t)$ and
matrix functions $S_k(t)$ given by (\ref{3.25}) and (\ref{3.11})
are invertible in the domain $-\ve_1<t < \ve_2$, $0 \leq k < K$.
Then the matrix functions $\wt C$ and $\wt Q$ are well defined by
(\ref{3.15})--(\ref{3.17}) and satisfy MTL in the domain $-\ve_1<t
< \ve_2$, $0 < k< K$.
\end{Tm}
\begin{proof}.
To prove the theorem we shall obtain derivatives in $t$
 for $\wh \Pi_k$, $S_k$, $\wh \Pi_k
S_k^{-1}$, and, finally, $\wh w_A(k, \l):=U(k)^{-1}w_A(k,
\l)U(k)$.  First let us recall (\ref{3.10}) and show by induction
that
\begin{equation} \label{3.18}
\frac{d}{d t} \wh \Pi_k(t)= A  \wh \Pi_{k}(t) {\clp }_2 +i \wh
\Pi_k(t) \psi(k,t).
\end{equation}
In view of (\ref{3.9}) we get
\begin{equation} \label{3.19}
 \wh \Pi^{\prime}_l= \wh \Pi^{\prime}_{l-1} \xi(l)^{-1}-i A \wh
\Pi^{\prime}_{l-1} {\clp }_2-  \wh \Pi_{l-1}
\xi(l)^{-1}\xi(l)^{\prime} \xi(l)^{-1}.
\end{equation}
Taking into account (\ref{3.2}) we  see that
\begin{equation} \label{3.21}
\xi(l)^{\prime}= \left[
\begin{array}{lr}
-i \big(C(l+1)C(l)^{-1}- C(l)C(l-1)^{-1} \big)  & Q(l) C(l)
\\ -C(l)^{-1} Q(l) & 0
\end{array}
\right].
\end{equation}
Suppose that (\ref{3.18}) is true for $k=l-1$. Then the definition
of $\psi$ and equalities (\ref{3.20}), (\ref{3.19}) yield
\[
\wh \Pi^{\prime}_l= \big( A  \wh \Pi_{l-1}{\clp }_2 +i \wh
\Pi_{l-1} \psi(l-1) \big) \xi(l)^{-1}-i A  \big( A  \wh
\Pi_{l-1}{\clp }_2 +i \wh \Pi_{l-1} \psi(l-1) \big) \clp_2
\] \[
-
 \wh \Pi_{l-1}
\xi(l)^{-1}\xi(l)^{\prime} \xi(l)^{-1}= A \big( -i A  \wh
\Pi_{l-1}{\clp }_2+ \wh \Pi_{l-1} \xi(l)^{-1} \clp_2 \big)
\]
\begin{equation} \label{3.22}
 + A \wh
\Pi_{l-1}\xi(l)^{-1}  \clp_1 + \wh \Pi_{l-1} \xi(l)^{-1} \big( i
\xi(l) \psi(l-1)- \xi(l)^{\prime} \big)\xi(l)^{-1}.
\end{equation}
 Using (\ref{3.20}) and (\ref{3.21}) we calculate
directly that
\begin{equation} \label{3.23}
\xi(l)^{-1} \clp_1= \clp_2 \psi(l), \quad \big( i \xi(l)
\psi(l-1)- \xi(l)^{\prime} \big)\xi(l)^{-1}=i \psi(l).
\end{equation}
From (\ref{3.9}), (\ref{3.22}) and (\ref{3.23}) follows equation
(\ref{3.18}) for $k=l$, i.e., (\ref{3.18}) holds for all $k \geq
0$.

Taking into account (\ref{3.11}) and (\ref{3.18}) one can prove
that
\begin{equation} \label{3.24}
\frac{d}{d t} S_k(t)=\frac{1}{2} \Big(AS_k(t)+S_k(t)A^*+ i \wh
\Pi_{k}(t) \left[
\begin{array}{lr}
0 & -I_p \\ I_p & 0
\end{array}
\right] \wh \Pi_{k}(t)^* \Big).
\end{equation}
Suppose (\ref{3.24}) is true for $k=l-1$. Then from formulas
(\ref{3.25})  and (\ref{3.18}) and equation (\ref{3.24}) for
$k=l-1$ we obtain
\[
S^{\prime}_l=\frac{1}{2} \Big( AS_{l-1}+S_{l-1}A^*+ i \wh
\Pi_{l-1} \left[
\begin{array}{lr}
0 & -I_p \\ I_p & 0
\end{array}
\right] \wh \Pi_{l-1}^* \Big) + \wh \Pi_{l-1} \zeta^{\prime}(l)
\wh \Pi_{l-1}^*
\]
\begin{equation} \label{3.26}
+ \big(  A  \wh \Pi_{l-1} {\clp }_2 +i \wh \Pi_{l-1} \psi(l-1)
\big) \zeta(l) \wh \Pi_{l-1}^*+  \wh \Pi_{l-1} \zeta(l) \big(  A
\wh \Pi_{l-1} {\clp }_2 +i \wh \Pi_{l-1} \psi(l-1) \big)^*.
\end{equation}
By (\ref{3.9}), (\ref{3.25}), and (\ref{3.27}) after some
calculations we derive now (\ref{3.24}) for $k=l$ and thus for all
$k \geq 0$.

By (\ref{3.18}) and (\ref{3.24}) one can show that
\begin{equation} \label{3.28}
\frac{d}{d t} \big( \wh \Pi_k(t)^* S_k(t)^{-1} \big)=- \clp_1 \wh
\Pi_k(t)^* S_k(t)^{-1}A-i J \wt \psi(k,t) J  \wh \Pi_k(t)^*
S_k(t)^{-1},
\end{equation}
\begin{equation} \label{3.29}
\wt \psi(k,t):=  \left[ \begin{array}{lr} 0 & \wt C(k+1,t)
 \\ \wt C(k, t)^{-1} & 0 \end{array} \right]
\end{equation}
where  matrix functions $\wt C(k,t)$ are given by (\ref{3.15}),
(\ref{3.17}). Indeed, by (\ref{3.18}) and (\ref{3.24}) we have
\[
\big( \wh \Pi_k^* S_k^{-1} \big)^{\prime}= \clp_2 \wh \Pi_k^* A^*
S_k^{-1}- i J \psi(k) J \wh \Pi_k^* S_k^{-1}
\]
\[
- \frac{1}{2} \wh \Pi_k^* S_k^{-1} \left( A S_k+ S_k A^*+ i \wh
\Pi_{k} \left[
\begin{array}{lr}
0 & -I_p \\ I_p & 0
\end{array}
\right] \wh \Pi_{k}^* \right) S_k^{-1}=- \frac{1}{2} \wh \Pi_k^*
S_k^{-1}A
\]
\begin{equation} \label{3.30}
- \frac{1}{2}(\clp_1- \clp_2) \wh \Pi_k^*A^* S_k^{-1}-i  J \psi(k)
J \wh \Pi_k^* S_k^{-1}+ \frac{i}{2}X(k)(\clp_1- \clp_2)J \wh
\Pi_k^* S_k^{-1}.
\end{equation}
Rewrite (\ref{3.14}) in the form
\begin{equation} \label{3.31}
A^* S_k^{-1}=S_k^{-1}A-i S_k^{-1}\wh \Pi_{k} J \wh \Pi_{k}^*
S_k^{-1}.
\end{equation}
Substitute  $A^* S_k^{-1}$  by the right-hand side of (\ref{3.31})
in the  right-hand side of (\ref{3.30}) to get
\[
\frac{d}{d t} \big( \wh \Pi_k(t)^* S_k(t)^{-1} \big)=- \clp_1 \wh
\Pi_k(t)^* S_k(t)^{-1}A
\] \[
-i \left[ \begin{array}{lr} 0 & C(k,t)^{-1}- X_{11}(k,t) \\ C(k+1,
t)+ X_{22}(k,t) & 0 \end{array} \right] \wh \Pi_k(t)^*
S_k(t)^{-1}.
\]
So according to (\ref{3.15}) it remains to obtain the equality
\begin{equation} \label{3.31.1}
\wt C(k,t)^{-1}=C(k,t)^{-1}- X_{11}(k,t)
\end{equation}
to prove (\ref{3.28}). By definition (\ref{3.15}) equality
(\ref{3.31.1}) is true for $k=0$. To prove (\ref{3.31.1}) for
$k>0$ notice that $\eta(k) J \xi(k)=\clp_1$ and in view of
(\ref{3.4})-(\ref{3.7}) derive the relation  ( \cite{SaL2},
formula (8.2.7)):
\begin{equation} \label{3.31.1'}
U(k,t) \clp_1 =J H_k(t)U(k-1,t).
\end{equation}
Putting
\begin{equation} \label{3.31.2}
\wt \xi(k,t):=\wt U(k,t)^{-1} \wt U(k-1,t), \quad \wt
U(k,t):=w_0(k,t)^{-1}U(k,t),
\end{equation}
consider the expression $\wt \xi$. By (\ref{2.7}) we have
\[
\wt \xi(k)=U(k)^{-1}w_0(k)w_0(k-1)^{-1}U(k-1)=
\]
\begin{equation} \label{3.31.2'}
U(k)^{-1}(I_m- J \Pi_k^* S_k^{-1} \Pi_k J H_k  + J H_k J
\Pi_{k-1}^* S_{k-1}^{-1} \Pi_{k-1})U(k-1).
\end{equation}
Hence taking into account formulas (\ref{3.7}), (\ref{3.31.1'})
and definitions (\ref{3.5}), (\ref{3.8}) and (\ref{3.17})
 we get
 \[
 \wt \xi(k)= \xi(k) - J X(k) \clp_1 +  \clp_1 J X(k-1)=
 \]
\begin{equation} \label{3.31.3}
\left[
\begin{array}{lr} -i \big(Q(k)+i (X_{21}(k-1)-X_{21}(k)) \big) & C(k)+
X_{22}(k-1) \\ C(k)^{-1}- X_{11}(k) & 0 \end{array} \right].
\end{equation}
By (\ref{3.31.3}) the left lower $p \times p$ block $( \wt \xi(k)J
\wt \xi(k)^*)_{21}$ of $\wt \xi(k)J \wt \xi(k)^*$ equals
$(C(k)^{-1}- X_{11}(k))(C(k)+ X_{22}(k-1) )$, and by
(\ref{3.31.2}) $\wt \xi$ is $J$-unitary. Thus we have
\[
(C(k)^{-1}- X_{11}(k))(C(k)+ X_{22}(k-1) )=I_p,
\]
i.e., equality (\ref{3.31.1}) holds. Moreover, from the
definitions (\ref{3.15}), (\ref{3.16}) and equalities
(\ref{3.31.1}) and (\ref{3.31.3}) it follows that
\begin{equation} \label{3.31.4}
 \wt \xi(k,t)= \left[
\begin{array}{lr} -i \wt Q(k,t) & \wt C(k) \\ \wt C(k)^{-1}
 & 0 \end{array} \right].
\end{equation}
Now again use the fact that $\xi$ is $J$-unitary to show that
(\ref{3.3}) for $\wt C$ and $\wt Q$ holds: $C(k)Q(k)^*=Q(k)C(k)$.

Next we shall calculate the derivative of
\begin{equation} \label{3.32}
 \wh w_A(k,t, \l)=
U(k,t)^{-1}w_A(k,t, \l)U(k,t).
\end{equation}
According to the definitions (\ref{2.6}) and (\ref{3.8}) the
function $\wh w_A$ has the form
\begin{equation} \label{3.33}
\wh w_A(k,t, \l)=I_m-i J \wh \Pi_k(t)^* S_k(t)^{-1}(A- \l
I_n)^{-1} \wh \Pi_k(t).
\end{equation}
Using formulas (\ref{3.18}) and (\ref{3.28}) and equality $A= (A-
\l I_n) + \l I_n$ we obtain easily:
\[
\frac{d}{d t} \wh w_A(k, \l)= \l \big( -J \clp_1 J (\wh w_A(k,
\l)-I_m)+ (\wh w_A(k, \l)-I_m) \clp_2 \big)-
\]
\begin{equation} \label{3.34}
i \wt \psi(k) (\wh w_A(k, \l)-I_m)+i (\wh w_A(k, \l)-I_m) \psi(k)
+i J \clp_1 X(k) - i J  X(k) \clp_2.
\end{equation}
Now in view of (\ref{3.15}), (\ref{3.29}) and (\ref{3.31.1}) after
evident transformations of (\ref{3.34}) we get the result
\begin{equation} \label{3.35}
\frac{d}{d t} \wh w_A(k,t, \l)= \wt F(k,t, \l)\wh w_A(k,t, \l)-\wh
w_A(k,t, \l)F(k,t, \l),
\end{equation}
where
\begin{equation} \label{3.36}
F(k,t, \l)=- \l \clp_2 -i \psi(k,t), \quad \wt F(k,t, \l)=- \l
\clp_2 -i \wt \psi(k,t).
\end{equation}

We shall need also a reformulation of (\ref{2.15}) that follows
from the relations (\ref{3.31.1'}), (\ref{3.31.2'}) and
(\ref{3.32}):
\begin{equation} \label{3.37}
\wh w_A(k,t, \l)G(k,t, \l)= \wt G(k,t, \l)\wh w_A(k-1,t, \l),
\end{equation}
where
\begin{equation} \label{3.38}
G(k,t, \l)=- \l \clp_1 +i \xi(k,t), \quad \wt G(k,t, \l)=- \l
\clp_1 +i \wt \xi(k,t).
\end{equation}
Recall that system (\ref{3.2}) is equivalent to the zero curvature
equation (\ref{3.39}). Equation (\ref{3.39}) is a compatibility
condition for the systems
\begin{equation} \label{3.40}
W(k,t, \l)=G(k,t, \l)W(k-1,t, \l), \quad \frac{d}{d t}W(k,t,
\l)=F(k,t, \l)W(k,t, \l).
\end{equation}
Thus by (\ref{3.2}) there exists an $m \times m$ matrix function
$W$ ($\det \, W \not= 0$) that satisfies (\ref{3.40}) \cite{SaL2}.
For $\wt W:= \wh w_A W $ formula (\ref{3.35}) and the second
relation in (\ref{3.40}) now yield:
\begin{equation} \label{3.41}
\frac{d}{d t} \wt W(k,t, \l)= \wt F(k,t, \l)\wt W(k,t, \l).
\end{equation}
Formula (\ref{3.37}) and the first relation in (\ref{3.40}) yield
\begin{equation} \label{3.42}
\wt W(k,t, \l)=\wh w_A(k,t, \l)G(k,t, \l)W(k-1,t, \l)= \wt G(k,t,
\l)\wt W(k-1,t, \l).
\end{equation}
In view of formula (\ref{3.31.4}) and the second relations in
formulas (\ref{3.36}) and (\ref{3.38}) the compatibility condition
\begin{equation} \label{3.42'}
\frac{d}{d t} \wt G(k,t, \l)+ \wt G(k,t, \l) \wt F(k-1,t, \l)- \wt
F(k,t, \l) \wt G(k,t, \l)=0
\end{equation}
for systems (\ref{3.41}) and (\ref{3.42}) is equivalent to the
system
\begin{equation} \label{3.43}
\displaystyle{\left\{
\begin{array}{c} \displaystyle{\wt Q^{\prime}(k,t)=
\wt C(k+1,t) \wt C(k,t)^{-1}-
\wt C(k,t) \wt C(k-1,t)^{-1}} \\ \displaystyle{\wt
C^{\prime}(k,t)= \wt Q(k,t) \wt C(k,t).}
\end{array}  \right.}
\end{equation}
Thus $\wt Q$ and $\wt C$ satisfy MTL.
\end{proof}
%%%%%%%%%%%%%%%%%%%%%%%%%%%%%%%%%%%%%%%%%%%%%%%%%%%%%%%
\section{Explicit solutions}
\label{ES} \setcounter{equation}{0} \vspace{-1mm}

In this section we shall consider first the case of the  trivial
MTL (\ref{3.2}), (\ref{3.3}) solution that is given by the
functions
\begin{equation} \label{j5.1}
C(k,t) \equiv I_p \quad (k \geq 0), \quad Q(k,t) \equiv 0 \quad (k
> 0).
\end{equation}
Starting from these initial $C$ and $Q$ we shall use GBDT to get
explicit formulas for a wide class of the MTL solutions,
fundamental DCS solutions and evolution of the Weyl functions.
According to (\ref{3.6}) we have
\begin{equation} \label{j5.2}
\psi(k,t) \equiv J \quad (k \geq 0), \quad \xi(k,t) \equiv J \quad
(k> 0).
\end{equation}
Putting $U(0,t) \equiv I_m$ we derive also
\begin{equation} \label{j5.3}
U(k,t)=J^k, \quad H_k(t)=J^{k} {\cal{P}}_2 J^k.
\end{equation}
Thus the normalized fundamental solution of the DCS has the form
\begin{equation} \label{j5.4}
w(k, \l)= \prod_{l=1}^k(I_m+ i \l J^{l+1}{\cal{P}}_2 J^l).
\end{equation}
Notice that ${\cal{P}}_1+{\cal{P}}_2=I_m$ and
$J{\cal{P}}_2J={\cal{P}}_1$. Hence we obtain
\begin{equation} \label{j5.5}
(I_m+i \l J {\cal{P}}_2)(I_m+i \l J {\cal{P}}_1)=I_m+i \l J - \l^2
{\cal{P}}_1= \left[
\begin{array}{lr}
(1- \l^2 )I_p & i \l I_p \\ i \l I_p & I_p
\end{array}
\right].
\end{equation}
From (\ref{j5.5}) follows representation
\begin{equation} \label{j5.6}
(I_m+i \l J {\cal{P}}_2)(I_m+i \l J {\cal{P}}_1)=TD_{\l}T^{-1},
\end{equation}
where
\begin{equation} \label{j5.7}
T= \left[
\begin{array}{lr}
I_p &  I_p \\ \frac{\l+ \sqrt{\l^2-4}}{2 i}I_p &  \frac{\l-
\sqrt{\l^2-4}}{2 i} I_p
\end{array}
\right], \quad D_{\l}= \left[
\begin{array}{lr}
z_1 I_p & 0 \\ 0 & z_2 I_p
\end{array}
\right],
\end{equation}
\begin{equation} \label{j5.8}
z_{1,2}= \frac{1}{2} (2- \l^2 \pm \l \sqrt{\l^2-4}).
\end{equation}
Later we shall choose $ \sqrt{\l^2-4} \in \BC_+$ for $\l \in
\BC_+$. As $J^2=I_m$, using (\ref{j5.4}) and (\ref{j5.6}) we
obtain:
\begin{equation} \label{j5.9}
w(k, \l)=TD_{\l}^lT^{-1} \, {\mathrm{for}} \, k=2l, \quad w(k,
\l)=(I_m+i \l J {\cal{P}}_1)TD_{\l}^lT^{-1} \, {\mathrm{for}} \,
k=2l+1.
\end{equation}
We shall apply now Theorem \ref{Tm3.1} to construct explicit MTL
solutions.
\begin{Tm} \label{Tmj5.1}
Fix $n>0$, two $n \times n$ parameter matrices $A$ and $S_0(0)$
and $n \times m $ parameter matrix $\wh \Pi_0(0)=[\t_1(0,0) \quad
\t_2(0,0)]$ with $n \times p$ blocks $\t_l(0,0)$, such that
\begin{equation} \label{j5.10}
A= \a+ \a^{-1}, \quad  \det (\a- \a^{-1}) \not=0, \quad S_0(0)>0,
\end{equation}
\begin{equation} \label{j5.11}
\det \left(I_p-[I_p \quad 0]\wh \Pi_0(0)^*S_0(0)^{-1}\wh \Pi_0(0)
\left[
\begin{array}{c} I_p \\  0
\end{array}
\right] \right) \not=0,
\end{equation}
and (\ref{3.12}) holds. Put now
\begin{equation} \label{j5.12}
\wt C(k,t)=I_p+X_{22}(k-1,t) \quad (k>0), \quad \wt C(0,t)= \big(
I_p-X_{11}(0,t) \big)^{-1},
\end{equation}
\begin{equation} \label{j5.13}
\wt Q(k,t)=i(X_{21}(k-1,t)-X_{21}(k,t))  \quad (k>0),
\end{equation}
where $X_{ij}(k,t)$ are $p \times p$ blocks of the matrix
\begin{equation} \label{j5.14}
X(k,t)=\left[
\begin{array}{lr}
X_{11}(k,t) & X_{12}(k,t) \\ X_{21}(k,t) & X_{22}(k,t)
\end{array}
\right]=\wh \Pi_k(t)^* S_k(t)^{-1} \wh \Pi_k(t).
\end{equation}
Define matrices $\wh \Pi_k$ and $S_k$ in the right-hand side of
(\ref{j5.14}) by the equalities
\begin{equation} \label{j5.16}
\left[
\begin{array}{c}
\t_+ \\ \t_-
\end{array}
\right]= \left( \left[
\begin{array}{lr}
i \a^{-1} & i \a \\ I_p & I_p
\end{array}
\right] \right)^{-1} \left[
\begin{array}{c}
\t_1(0,0) \\ \t_2(0,0)
\end{array}
\right],
\end{equation}
\[
\wh \Pi_k(t)=[\t_1(k,t) \quad \t_2(k,t)], \quad \t_1(k,t)= (-i
\a)^{k-1}e^{\a t} \t_+ +(-i \a^{-1})^{k-1}e^{\a^{-1}t} \t_-,
\] \begin{equation} \label{j5.15}
 \t_2(k,t)= (-i
\a)^{k}e^{\a t} \t_+ +(-i \a^{-1})^{k}e^{\a^{-1}t} \t_-,
\end{equation}
\begin{equation} \label{j5.17}
S_k(t)=S_0(t)+ \sum_{l=0}^{k-1} \t_2(l,t) \t_2(l,t)^*, \quad
S_0(t)= \frac{1}{2} \t_1(0,t) \t_1(0,t)^*+e^{ \frac{1}{2}
At}q(t)e^{ \frac{1}{2} A^*t},
\end{equation}
\[
q(t)=S_0(0)-\frac{1}{2} \t_1(0,0) \t_1(0,0)^* +
\]
\begin{equation} \label{j5.18}
\frac{1}{4} \int_0^t e^{ -\frac{1}{2} A u}(A \t_1(0,u)
\t_1(0,u)^*+ \t_1(0,u) \t_1(0,u)^*A^*) e^{- \frac{1}{2} A^*u}d u.
\end{equation}
Then for some $\varepsilon_1 >0$, $\ve_2>0$ matrix functions $\wt
C$ and $\wt Q$ are well defined by (\ref{j5.12})--(\ref{j5.18})
and satisfy MTL in the domain $-\ve_1<t < \ve_2$, $0 < k< \infty$.
\end{Tm}
\begin{proof}.
One can check easily that  according to the second relation in
(\ref{j5.10}) the square  matrix in the right-hand side of
(\ref{j5.16}) is invertible. By  (\ref{j5.16}) the right-hand
sides in the second and third relations in (\ref{j5.15}) at $k=0$,
$t=0$ turn into $\t_1(0,0)$ and $\t_2(0,0)$, respectively, i.e.,
$\wh \Pi$ is well defined by (\ref{j5.15}). Moreover, from
(\ref{j5.15}) we get $\t_1'(0,t)=i \t_2(0,t)$ and
$\t_2'(0,t)=A\t_2(0,t)+i \t_1(0,t)$. Hence in view of (\ref{j5.2})
$\wh \Pi_0(t)$ satisfies (\ref{3.10}). Notice also that
\begin{equation} \label{j5.19}
\big(\t_1(0,t) \t_1(0,t)^* \big)'=i\big(\t_2(0,t) \t_1(0,t)^*
-\t_1(0,t) \t_2(0,t)^*\big).
\end{equation}
From the equalities $\t_1(k,t)=\t_2(k-1,t)$ and
$\t_2(k,t)=\t_1(k-1,t)-i A \t_2(k-1,t)$ it follows that $\wh
\Pi_k(t)$ satisfies (\ref{3.9}). According to (\ref{j5.17}) we see
that $S_k(t)$ satisfies (\ref{3.25}). Finally formulas
(\ref{j5.17}), (\ref{j5.18}) yield
\begin{equation} \label{j5.20}
\big(S_0(t)-\frac{1}{2} \t_1(0,t) \t_1(0,t)^* \big)'= \frac{1}{2}
(AS_0(t)+S_0(t)A^*),
\end{equation}
and taking into account (\ref{j5.19}) we obtain (\ref{3.11}).
 Therefore
matrices $\wt C$ and $\wt Q$ coincide with the matrices generated
in Theorem \ref{Tm3.1}, when $C$ and $Q$ are given by
(\ref{j5.1}). As $S_0(0)>0$ and inequality (\ref{j5.11}) holds, so
for some $\varepsilon_1
>0$, $\ve_2>0$ we have $S_0(t)>0$, $\det(I_p-X_{11}(0,t)) \not= 0$
on the interval $-\ve_1<t < \ve_2$. Taking into account the first
equality in (\ref{j5.17}) we derive $S_k(t)>0$ also.  The
conditions of Theorem \ref{Tm3.1} are fulfilled, i.e., $\wt C$ and
$\wt Q$ satisfy MTL.
\end{proof}
Introduce matrix functions $\wt H_k(t)$ and $\wt U_k(t)$ by the
equalities
\begin{equation} \label{j5.21}
\wt H_k(t)=J \wt U(k,t)\wt \eta(k,t) \wt U(k,t)^* J, \quad \wt
U(k,t)=\wt U(k-1,t) \wt \xi(k,t)^{-1} ,
\end{equation}
where $U(0,t)=I_m$,
\begin{equation} \label{j5.21'}
\wt \eta(k,t)= \left[
\begin{array}{lr}
\wt C(k,t) & 0 \\ 0 & 0
\end{array}
\right], \quad
 \wt \xi(k,t)= \left[
\begin{array}{lr} -i \wt Q(k,t) & \wt C(k) \\ \wt C(k)^{-1}
 & 0 \end{array} \right].
\end{equation}
\begin{Tm} \label{Tmj5.2} Suppose the conditions of Theorem
\ref{Tmj5.1} are fulfilled. Then the fundamental solutions of the
DCS (\ref{2.9}), where $\wt H_k(t)$ is given by (\ref{j5.21}), in
the case $k=2l$ take the form
\begin{equation} \label{j5.22}
\wt w(k,t, \l)= \wt U(k,t) \wh w_A(k,t, \l)TD_{\l}^lT^{-1} \wh
w_A(0,t, \l)^{-1} ,
\end{equation}
 and in the case $k=2l+1$ take the form
\begin{equation} \label{j5.23}
\wt w(k,t, \l)=\wt U(k,t) \wh w_A(k,t, \l)J(I_m+i \l J
{\cal{P}}_1)TD_{\l}^lT^{-1}  \wh w_A(0,t, \l)^{-1}  .
\end{equation}
Here $\wh w_A$ is defined in (\ref{3.33}), matrix functions $\wt
C$, $\wt Q$, $\wh \Pi_k$, and $S_k$ are defined in
(\ref{j5.12})-(\ref{j5.18}). If $\det \, A \not=0$, then we have
\begin{equation} \label{j5.26}
\wt U(k,t)=J^k \wh w_A(k,t, 0) J^k  \wh w_A(0,t, 0)^{-1}.
\end{equation}
The evolution of the corresponding Weyl function for DCS on the
semiaxis is given by the formula
\begin{equation} \label{j5.24}
\wt \varphi(t, \l)= i[0 \quad I_p] \wh w_A(0,t, \l) h(\l) \Big(
[I_p \quad 0] \wh w_A(0,t, \l) h(\l) \Big)^{-1},
\end{equation}
where
\begin{equation} \label{j5.25}
h(\l)= \left[
\begin{array}{c}
I_p \\  \frac{\l+ \sqrt{\l^2-4}}{2 i}I_p
\end{array}
\right] \quad (\l \in \BC_+).
\end{equation}
\end{Tm}
\begin{proof}.
According to the first relation in (\ref{3.31.2}) and formula
(\ref{3.31.4}) our matrix functions $\wt U$ and $\wt \xi$ given by
the formulas (\ref{j5.21}) and (\ref{j5.21'}) coincide with those
treated in Section \ref{TC}. Recall that by (\ref{2.24'}) we have
$\wt H_k=Jw_0(k)^{-1}JH_k w_0(k-1)$. Hence in view of formula
(\ref{3.31.1'}) and the second relation in formula (\ref{3.31.2})
we obtain $\wt H_k=J \wt U(k) {\cal P}_1 \wt U(k-1)^{-1}$. Notice
that $\wt U$ is $J$-unitary. So according to the first relation in
(\ref{3.31.2}) we get
\begin{equation} \label{j5.27}
\wt H_k(t)=J \wt U(k,t) {\cal P}_1 \wt \xi(k,t)^{-1} \wt
U(k,t)^{-1}=J \wt U(k,t) {\cal P}_1 \wt \xi(k,t)^{-1} J \wt
U(k,t)^{*}J.
\end{equation}
From (\ref{j5.21'}) it follows that ${\cal P}_1 \wt \xi^{-1} J=
\wt \eta$, i.e., (\ref{j5.27}) is equivalent to the first relation
in (\ref{j5.21}). In other words our $\wt H_k$ coincides with $\wt
H_k$ treated in Section \ref{GBDT} and we can apply Theorem
\ref{Tm2.1}. Therefore formulas (\ref{2.8}), (\ref{3.31.2}), and
(\ref{3.32}), and normalizing condition $\wt w(0,t, \l)=I_m$
yield:
\[
\wt w(k,t, \l)=\wt U(k,t) \wh w_A(k,t, \l)U(k,t)^{-1}w(k,t, \l)
\wh w_A(0,t, \l)^{-1}  .
\]
Taking into account (\ref{j5.3}) and (\ref{j5.9}) we derive now
(\ref{j5.22}) and (\ref{j5.23}). Representation (\ref{j5.26})
follows from Remark \ref{Rk2.2}.

To derive (\ref{j5.24}) we first need to  show that $\vp (\l):=
\frac{1}{2} \big( \l+ \sqrt{\l^2-4} \big)I_p$ is a Weyl function
of the initial DCS. Indeed, choosing $K=2l$, by (\ref{j5.7}) and
(\ref{j5.9}) we see that $w(K,\l)h(\l)=z_1^l h(\l)$. Notice that
$i(\vp (\l)^*- \vp(\l))>0$ $(\l \in \BC_+)$, i.e., $h(\l)^*J
h(\l)>0$. From $w(K,\l)h(\l)=z_1^l h(\l)$ and $h(\l)^*J h(\l)>0$,
using (\ref{2.27}), we obtain the inequality
\[
h(\l)^*\sum_{k=0}^{K-1}w(k, \l)^*H_{k+1} w(k, \l)h(\l)=
\]
\begin{equation} \label{j5.28}
\frac{i(1-|z_1|^K)}{ \l - \overline{\l}}h(\l)^*J h(\l) < \frac{i}{
\l - \overline{\l}}h(\l)^*J h(\l).
\end{equation}
By (\ref{j5.28}) the inequality (\ref{2.29}) for $\vp (\l)=
\frac{1}{2} \big( \l+ \sqrt{\l^2-4} \big)I_p$ is valid, and thus
$\vp$ is a Weyl function for the initial DCS with $H_k=J^k{\cal
P}_2J^k$ on the semiaxis. We apply now Theorem \ref{Tm2.5} and it
remains to prove that
\begin{equation} \label{j5.29}
\det \, [I_p \quad 0] \wh w_A(0,t, \l)h(\l) \not=0.
\end{equation}
Recall that $h(\l)^*J h(\l)>0$ and $\wh w_A(0,t, \l)^*J\wh
w_A(0,t, \l) \geq J$ for $\l \in \BC_+$, $\l \not\in \s(A)$.
Therefore we have $\wt h(\l)^*J \wt h(\l)>0$ for $\wt h(\l)=\wh
w_A(0,t, \l)h(\l)$. Suppose there is a vector $f \not=0$ such that
$[I_p \quad 0] \wt h(\mu)f =0$ for some $\mu \in \BC_+$, $\mu
\not\in \s(A)$. Then, taking into account inequality $f^* \wt
h(\mu)^*J \wt h(\mu)f>0$, we see that the $(p+1)$-dimensional
subspace $L_f= {\mathrm{span}} \left(h(\mu)f
\cup{\mathrm{Im}}\left[
\begin{array}{c}
  0 \\ I_p
\end{array}
\right] \right)$ is $J$-nonnegative. (Here Im means image and
$L_f$ is called $J$-nonnegative if for any $g \in L_f$ we have
$g^*J g \geq 0$.) As the maximal dimension of the $J$-nonnegative
subspaces is $p$, so we come to a contradiction, i.e., inequality
(\ref{j5.29}) is true. Therefore by Theorem \ref{Tm2.5} the last
statement of the theorem is true also.
\end{proof}
\begin{equation} \label{j5.30}
\end{equation}
%%%%%%%%%%%%%%%%%%%%%%%%%%%%%%%%%%%%%%%%%%%%%%%%%%%%%%%%%
\section{Non-Abelian Toda lattice }
\label{NTL} \setcounter{equation}{0} \vspace{-1mm}

To describe a B\"acklund-Darboux transformation for the NTL
(\ref{3.2}) we introduce $\wh \Pi_k(t)$ by the equations
(\ref{3.9}) and (\ref{3.10}) as before, and introduce an $n \times
n$ matrix $B$ and a new set of $m \times n$ matrix functions
$\L_k(t)$. These matrix functions are given by the initial value
$\L_0(0)$ and equations
\begin{equation} \label{4.5}
\L_k(t)=  \xi(k,t) \L_{k-1}(t)+i{\clp }_1 \L_{k-1}(t) B,
\end{equation}
\begin{equation} \label{4.6}
\frac{d}{d t}  \L_0(t)=  {\clp }_1 \L_0(t)B -i\psi(0,t) \L_0(t).
\end{equation}
Under the previous section assumptions we had $B=A^*$, $\L_k=iJ
\wh \Pi_k^*$. In this section we have five parameter matrices:
$A$, $B$, $\wh \Pi_0(0)$, $\L_0(0)$ and $S_0(0)$ and assume that
the operator identity
\begin{equation} \label{4.7}
AS_0(0)-S_0(0)B= \wh \Pi_0(0) \L_0(0)
\end{equation}
holds. Now the matrix functions $S_k(t)$ are defined by the
relations
\begin{equation} \label{4.8}
S_k(t)=S_{k-1}(t)-i \wh \Pi_{k-1}(t) \clp_2 \xi(k,t) \L_{k-1}(t),
\end{equation}
\begin{equation} \label{4.9}
\frac{d}{d t} S_0(t)=\frac{1}{2} \Big(AS_0(t)+S_0(t)B- \wh
\Pi_{0}(t)(\clp_1 - \clp_2) \L_{0}(t) \Big).
\end{equation}
Then quite similar to the formulas  (\ref{3.16}), (\ref{3.18}),
and (\ref{3.24}) we derive
\begin{equation} \label{4.10}
AS_k(t)-S_k(t)B= \wh \Pi_k(t) \L_k(t),
\end{equation}
\begin{equation} \label{4.11}
\frac{d}{d t}  \L_k(t)=  {\clp }_1 \L_k(t)B -i\psi(k,t) \L_k(t),
\end{equation}
\begin{equation} \label{4.12}
\frac{d}{d t} S_k(t)=\frac{1}{2} \Big(AS_k(t)+S_k(t)B- \wh
\Pi_{k}(t)(\clp_1 - \clp_2) \L_{k}(t) \Big).
\end{equation}
Finally put
\begin{equation} \label{4.13}
\wt C(k,t)= C(k,t)-i Z_{12}(k-1,t) \quad (k>0),
\end{equation}
\begin{equation} \label{4.14}
\wt C(0,t)= (C(0,t)^{-1}+i Z_{21}(0,t))^{-1},
\end{equation}
\begin{equation} \label{4.15}
\wt Q(k,t)= Q(k,t)+ Z_{11}(k-1,t)-Z_{11}(k,t) \quad (k>0),
\end{equation}
where $Z_{ij}(k)$ are $p \times p$ blocks given by the formulas
\begin{equation} \label{4.15'}
Z(k)=\{Z_{ij}(k) \}_{i,j=1}^2:=\L_k S_k^{-1} \wh \Pi_k.
\end{equation}
 The B\"acklund-Darboux transformation result is given
by the following generalization of Theorem \ref{Tm3.1}.
\begin{Tm} \label{Tm4.1}
Suppose matrix functions $C(k,t)$ and $Q(k,t)$ satisfy NTL
(\ref{3.2}) and $C(0,t)$ is continuous. Suppose additionally that
matrix functions $C(0,t)^{-1}+i Z_{21}(0,t)$ and matrix functions
$S_k(t)$ given by (\ref{4.8}), (\ref{4.9}), and (\ref{4.15'}) are
invertible in the domain $-\ve_1<t < \ve_2$, $0 \leq k < K$. Then
the matrix functions $\wt C$ and $\wt Q$ are well defined by
(\ref{4.13})--(\ref{4.15}) and satisfy NTL in the domain $-\ve_1<t
< \ve_2$, $0 < k< K$.
\end{Tm}
\begin{proof}.
The proof is similar to the proof of Theorem \ref{Tm3.1}. We shall
need the evident identities:
\begin{equation} \label{4.16}
\clp_2 \xi(k)= \xi(k)^{-1} \clp_1= \clp_2 \xi(k) \clp_1, \quad
\xi(k)^{-1} \clp_1 \xi(k) \clp_2=\clp_2 \xi(k)^2 \clp_2= \clp_2.
\end{equation}
 From (\ref{4.8}) it follows that
\begin{equation} \label{4.17}
S_k^{-1}=S_{k-1}^{-1}+i S_k^{-1} \wh \Pi_{k-1} \clp_2 \xi(k)
\L_{k-1} S_{k-1}^{-1}.
\end{equation}
Using (\ref{4.5}), (\ref{4.17}), and the first equality in
relations (\ref{4.16}) by induction we can show that
\begin{equation} \label{4.18}
\L_k(t) S_k(t)^{-1}= \wt \xi(k,t) \L_{k-1}(t) S_{k-1}(t)^{-1} +i
\clp_1 \L_{k-1}(t) S_{k-1}(t)^{-1} A,
\end{equation}
where
\begin{equation} \label{4.19}
\wt \xi(k,t): =  \xi(k,t)-i \clp_1 Z(k-1,t)+i Z(k,t) \clp_1.
\end{equation}
We cannot use $J$-unitary matrices to derive the equality
\begin{equation} \label{4.20}
\wt C(k,t)^{-1}=C(k,t)^{-1} +i Z_{21}(k,t),
\end{equation}
which is the  analog of the important formula (\ref{3.31.1}),
anymore. Therefore we shall consider now $\clp_2 \wt \xi(k)^2
\clp_2$ for $k>0$. (If $k=0$, then formula (\ref{4.20}) follows
from (\ref{4.14}).) By (\ref{4.16}) and (\ref{4.19}) we derive
\begin{equation} \label{4.21}
\clp_2 \wt \xi(k)^2 \clp_2 = \clp_2 \wt \xi(k) \clp_1(\xi(k)-i
Z(k-1)) \clp_2.
\end{equation}
Notice further that according to (\ref{3.9}) and (\ref{4.16}) we
have
\begin{equation} \label{4.22}
\clp_2 Z(k) \clp_1 \xi(k) \clp_2= \clp_2 \L_k S_k^{-1} \wh
\Pi_{k-1} \clp_2.
\end{equation}
Hence taking into account formula (\ref{4.18}) and equality
$\clp_2 \wt \xi(k)= \clp_2 \wt \xi(k) \clp_1$ we get
\begin{equation} \label{4.23}
\clp_2 Z(k) \clp_1 \xi(k) \clp_2= \clp_2 \wt \xi(k) \clp_1 Z(k-1)
\clp_2.
\end{equation}
Using equalities (\ref{4.16}) and (\ref{4.19}) by (\ref{4.21}) and
(\ref{4.23}) we obtain
\begin{equation} \label{4.24}
\clp_2 \wt \xi(k)^2 \clp_2= \clp_2 \wt \xi(k) \clp_1 \xi(k)
\clp_2-i\clp_2 Z(k) \clp_1 \xi(k) \clp_2= \clp_2.
\end{equation}
In view of the formulas (\ref{4.13}) and (\ref{4.19}) equality
(\ref{4.24}) yields (\ref{4.20}). Moreover $\wt \xi$ preserves the
form (\ref{3.31.4}) from the previous section.

The transfer matrix function $\wh w_A$ is given here by the
formula
\begin{equation} \label{4.25}
\wh w_A(k,t, \l)=I_m- \L_k(t)^* S_k(t)^{-1}(A- \l I_n)^{-1} \wh
\Pi_k(t).
\end{equation}
From (\ref{3.39'}), (\ref{3.9}), (\ref{4.16}), and (\ref{4.25}) it
follows that
\[
\wh w_A(k, \l)G(k, \l)=- \l \clp_1+i \xi(k)-i \L_kS_k^{-1}(A- \l
I_n)^{-1} \wh \Pi_{k-1}
\]
\begin{equation} \label{4.26}
-\L_kS_k^{-1} \wh \Pi_{k-1} \xi(k)^{-1} \clp_1.
\end{equation}
Taking into account (\ref{4.18}) we rewrite (\ref{4.26}):
\[
\wh w_A(k, \l)G(k, \l)=- \l \clp_1+i \xi(k)-Z(k) \clp_1+ i \wt
\xi(k)(\wh w_A(k-1, \l)-I_m)
\]
\begin{equation} \label{4.27}
+ \clp_1 \L_{k-1}S_{k-1}^{-1}A(A- \l I_n)^{-1} \wh \Pi_{k-1}.
\end{equation}
Finally using the equality $A=(A- \l I_n)+ \l I_n$ and definitions
(\ref{3.38}) and (\ref{4.19}) of $\wt G$ and $\wt \xi$,
respectively, we obtain
\begin{equation} \label{4.28}
\wh w_A(k,t, \l)G(k,t, \l)= \wt G(k,t, \l)\wh w_A(k-1,t, \l).
\end{equation}

The proof of (\ref{3.35}), where $F$ and $\wt F$ are defined by
(\ref{3.36}), and $\wt \psi$ is defined by (\ref{3.29}),
(\ref{4.13}) and (\ref{4.14}), is quite analogous to the one in
Section 3. After we recall that NTL is equivalent to the zero
curvature equation (\ref{3.39}) it remains to derive from
(\ref{3.35}) and (\ref{4.28}) the equation (\ref{3.42'}) and then
(\ref{3.43}). These arguments coincide with the corresponding
arguments from Theorem \ref{Tm3.1}.
 \end{proof}

Finally notice that the important case of blow-up of the NTL
solution can be studied in terms of the invertibility of $S_k(t)$.
\section{Appendix: Jacobi matrices }
\label{Ap} \setcounter{equation}{0} \vspace{-1mm}

DCS are closely related to the well-known Jacobi matrices. For
symmetric block Jacobi matrices we shall obtain an analog of
Theorem \ref{TmIns}. Suppose that the sets of matrices $\{C(k)
\}_{k \geq 0}$ and $\{Q(k) \}_{k > 0}$ such that
\begin{equation} \label{A-3}
C(k)Q(k)^*=Q(k)C(k) \quad (k>0) , \quad C(k)>0 \quad (k \geq 0),
\end{equation}
are given. Fix $n>0$, two $n \times n$ parameter matrices $A$ and
$S_0$ and $n \times m $ parameter matrix $\wh \Pi_0$, such that
\begin{equation} \label{A-2}
AS_0-S_0A^*=i \wh \Pi_0 J \wh \Pi_0^*, \quad S_0>0.
\end{equation}
Introduce matrices $\wh \Pi_k$, $S_k$, $\wt C(k)$, and $\wt Q(k)$
by the same equalities as in Section \ref{TC}  but without
dependence on $t$:
\begin{equation} \label{A-1}
\wh \Pi_k= \wh \Pi_{k-1} \xi(k)^{-1}-i A  \wh \Pi_{k-1} {\clp }_2,
\quad S_k=S_{k-1}+ \wh \Pi_{k-1} \zeta(k)  \wh \Pi_{k-1}^*,
\end{equation}
\begin{equation} \label{A}
\xi(k)= \left[
\begin{array}{lr}
-i Q(k) & C(k) \\  C(k)^{-1} & 0
\end{array}
\right], \quad \zeta(k)=\left[
\begin{array}{lr}
0 & 0 \\ 0 & C(k)^{-1}
\end{array}
\right],
\end{equation}
\begin{equation} \label{A0}
\wt C(k)=C(k)+X_{22}(k-1) \quad (k>0), \quad \wt C(0,t)= \big(
C(0)-X_{11}(0) \big)^{-1},
\end{equation}
\begin{equation} \label{A0'}
\wt Q(k)=Q(k)+i(X_{21}(k-1)-X_{21}(k)),  \quad X(k)= \{ X_{ij}
\}_{i,j=1}^2=\wh \Pi_k^* S_k^{-1} \wh \Pi_k.
\end{equation}
Introduce now a block Jacobi matrix by the equalities
\begin{equation} \label{A1}
\wt L=\left[
\begin{array}{cccccc}
\wt b_1 & \wt a_1 & 0 & 0 & 0 & \ldots \\ \wt c_2 & \wt b_2 & \wt
a_2 & 0 & 0 & \ldots \\  0 & \wt c_3 & \wt b_3 & \wt a_3 & 0 &
\ldots
\\ \ldots & \ldots & \ldots & \ldots & \ldots & \ldots
\end{array}
\right],
\end{equation}
where for $k>0$ we put
\begin{equation} \label{A2}
\wt a_k=-i \wt C(k)^{- \frac{1}{2}}\wt C(k+1)^{\frac{1}{2}}, \quad
\wt b_k= \wt b_k^*=\wt C(k)^{- \frac{1}{2}} \wt Q(k) \wt
C(k)^{\frac{1}{2}}, \quad \wt c_k= \wt a_{k-1}^*.
\end{equation}
Matrix $\wt L$ can be considered as the GBDT of the Jacobi matrix
$L$ defined  by the same formulas (\ref{A1}) and (\ref{A2}) but
with tildes  removed.
\begin{Tm} \label{TmA}
Suppose Jacobi matrix $\wt L$ is given by the formulas
(\ref{A-1})-(\ref{A2}) and relations (\ref{A-3}) and (\ref{A-2})
are valid. If   $f \not=0$ is an eigenvector of $A$ such that
\begin{equation} \label{A3}
A f= \mu f, \quad [I_p \quad 0] \wh \Pi_0^*S_0^{-1}f=0,
\end{equation}
then the vector $Y = {\mathrm{col}} \, [y_1 \, y_2 \, \ldots ]$,
where ${\mathrm{col}}$ means column and
\begin{equation} \label{A4}
y_k=[0 \quad \wt C(k)^{-\frac{1}{2}}] \wh \Pi_{k-1}^*
S_{k-1}^{-1}f,
\end{equation}
is an eigenvector of $\wt L$:
\begin{equation} \label{A5}
\wt L Y= \mu Y, \quad \sum_{k=1}^{\infty}y_k^*y_k < \infty .
\end{equation}
\end{Tm}
\begin{proof}.
First notice that the equality  $\wt b_k= \wt b_k^*$ in (\ref{A2})
follows from (\ref{A-3}). Next similar to the proof of the
equation (\ref{4.18}) we shall show that
\begin{equation} \label{A6}
\wh \Pi_k^* S_k^{-1}=i {\cal P}_2 \wh \Pi_{k-1}^* S_{k-1}^{-1}A +
\left[ \begin{array}{lr} 0 & \wt C(k)^{-1} \\ \wt C(k) & -i \wt
Q(k)
\end{array}
\right] \wh \Pi_{k-1}^* S_{k-1}^{-1}.
\end{equation}
Indeed, by (\ref{1.2}), (\ref{2.2}), (\ref{2.14}), and (\ref{3.8})
we get
\[
\wh \Pi_k^* S_k^{-1}=i U(k)^*H_k J (U(k-1)^*)^{-1}\wh \Pi_{k-1}^*
S_{k-1}^{-1}A+U(k)^* \big( I_m+
\] \begin{equation} \label{A7}
H_kJ \Pi_{k-1}^*S_{k-1}^{-1} \Pi_{k-1}J- \Pi_{k}^*S_{k}^{-1}
\Pi_{k}J H_k J \big) (U(k-1)^*)^{-1}\wh \Pi_{k-1}^* S_{k-1}^{-1}.
\end{equation}
By (\ref{3.4})-(\ref{3.7}) we rewrite (\ref{A7}) as
\[
\wh \Pi_k^* S_k^{-1}= i J \eta(k) (\xi(k)^{-1})^*\wh \Pi_{k-1}^*
S_{k-1}^{-1}A+ \big((\xi(k)^{-1})^*+
\] \begin{equation} \label{A8}
J \eta(k) (\xi(k)^{-1})^* \wh \Pi_{k-1}^*S_{k-1}^{-1} \wh
\Pi_{k-1}J- \wh \Pi_{k}^*S_{k}^{-1} \wh \Pi_{k} \eta(k)
(\xi(k)^{-1})^*\big) \wh \Pi_{k-1}^* S_{k-1}^{-1}.
\end{equation}
As $J \eta(k) (\xi(k)^{-1})^*={\cal P}_2$, so (\ref{A8}) yields
\[
\wh \Pi_k^* S_k^{-1}=i {\cal P}_2 \wh \Pi_{k-1}^* S_{k-1}^{-1}A +
\] \begin{equation} \label{A9}
\left[ \begin{array}{lr} 0 &  C(k)^{-1} \\  C(k) & -i C(k)  Q(k)^*
C(k)^{-1} \end{array} \right]+\left[ \begin{array}{lr} 0 &
-X_{11}(k) \\ X_{22}(k-1)  & X_{21}(k-1)-X_{21}(k)
\end{array}
\right].
\end{equation}
Finally, using (\ref{3.31.1}), (\ref{A-3}), (\ref{A})-(\ref{A0'}),
we transform (\ref{A9}) into (\ref{A6}).

Now, according to the first relations in (\ref{A2}) and (\ref{A3})
and to the formulas (\ref{A4}) and (\ref{A6}) we obtain
\[
\wt a_k y_{k+1}=-i \wt C(k)^{-\frac{1}{2}}[0 \quad I_p ] \wh
\Pi_{k}^* S_{k}^{-1}f=
\] \begin{equation} \label{A10}
\mu y_k -\wt C(k)^{- \frac{1}{2}} \wt Q(k) \wt
C(k)^{\frac{1}{2}}y_k-i  \wt C(k)^{\frac{1}{2}}[I_p \quad 0 ] \wh
\Pi_{k-1}^* S_{k-1}^{-1}f.
\end{equation}
Using the second relation in the definition (\ref{A2}) we rewrite
(\ref{A10}) as
\begin{equation} \label{A11}
\wt b_k y_k+ \wt a_k y_{k+1}= \mu y_k -i  \wt
C(k)^{\frac{1}{2}}[I_p \quad 0 ] \wh \Pi_{k-1}^* S_{k-1}^{-1}f.
\end{equation}
In particular, in view of the second relation in (\ref{A3})
formula (\ref{A11}) yields
\begin{equation} \label{A12}
\wt b_1 y_1+ \wt a_1 y_{2}= \mu y_1.
\end{equation}
For $k>1$ from (\ref{A6}) we derive
\[
i  \wt C(k)^{\frac{1}{2}}[I_p \quad 0 ] \wh \Pi_{k-1}^*
S_{k-1}^{-1}f=
\] \begin{equation} \label{A13}
i  \wt C(k)^{\frac{1}{2}}\wt C(k-1)^{-\frac{1}{2}}[0 \quad \wt
C(k-1)^{- \frac{1}{2}} ] \wh \Pi_{k-2}^* S_{k-2}^{-1}f= \wt
a_{k-1}^* y_{k-1}.
\end{equation}
By (\ref{A11}) and (\ref{A13}) we get
\begin{equation} \label{A14}
\wt c_k y_{k-1}+ \wt b_k y_k+ \wt a_k y_{k+1}= \mu y_k \quad
(k>1).
\end{equation}
Formulas (\ref{A12}) and (\ref{A14}) yield the first relation in
(\ref{A5}).

To prove the second relation in (\ref{A5}) consider expressions
$y_k^*y_k$. One can see that
\begin{equation} \label{A15}
y_k^*y_k=f^*S_{k-1}^{-1} \wh \Pi_{k-1} \wt \zeta (k) \wh
\Pi_{k-1}^*S_{k-1}^{-1}f, \quad \wt \zeta(k)=\left[
\begin{array}{lr}
0 & 0 \\ 0 & \wt C(k)^{-1}
\end{array}
\right].
\end{equation}
By  (\ref{3.8})  we rewrite (\ref{A15}) in the form
\begin{equation} \label{A16}
y_k^*y_k=g_{k-1}^*w_0(k-1)^*J U(k-1) \wt \zeta (k) U(k-1)^* J
w_0(k-1)g_{k-1},
\end{equation}
where $g_k$ are defined in (\ref{j1}). Hence, using (\ref{3.31.2})
and  the fact that $w_0(k)$ is $J$-unitary, we obtain
\begin{equation} \label{A17}
y_k^*y_k==g_{k-1}^*J \wt U(k-1) \wt \zeta (k) \wt U(k-1)^* J
g_{k-1}.
\end{equation}
As $\wt \xi (k) \wt \zeta (k) \wt \xi (k)^*= \wt \eta(k)$,
formulas (\ref{j5.21}) and (\ref{A17}) yield:
\begin{equation} \label{A18}
y_k^*y_k==g_{k-1}^*J \wt U(k) \wt \xi (k) \wt \zeta (k) \wt \xi
(k)^* \wt U(k)^* J g_{k-1}=g_{k-1}^* \wt H_k g_{k-1}.
\end{equation}
It was shown in Theorem \ref{TmIns} that
$\sum_{k=1}^{\infty}g_{k-1}^* \wt H_k g_{k-1}< \infty$, i.e., by
(\ref{A18}) the second relation in (\ref{A5}) is true.
\end{proof}

\end{document}